\documentclass[iop]{emulateapj}
\usepackage{natbib}
\usepackage{hyperref}
\usepackage{amsmath}
\usepackage{amssymb}
\usepackage{graphicx, graphics}
\usepackage{aas_macros}

\newcommand{\hmpc}{$h^{-1}{\rm Mpc}$}
\newcommand{\lbox}{L_{\rm box}}

\newcommand{\rhobar}{\bar{\rho}}
\newcommand{\pbar}{\bar{P}}
\newcommand{\phat}{\hat{P}} 
\newcommand{\Nki}{N_{k_i}}  

\newcommand{\Nkq}{N_{k_q}}   
\newcommand{\kv}{\vec{k}} 
\newcommand{\dk}{\delta_{\kv_i}}

\newcommand{\kthresh}{k_{\rm thresh}}

\newcommand{\xv}{\mathbf{x}} 
 
\newcommand{\dvec}{\mathbf{d}}

\slugcomment{LLNL-JRNL-471523}

\shorttitle{Large-scale mode resampling in $N$-body simulations}
\shortauthors{Schneider et al.}

\begin{document}

\title{Fast generation of ensembles of \\cosmological $N$-body simulations
via mode-resampling}

\author{Michael D. Schneider}
\affil{Lawrence Livermore National Laboratory, P.O. Box 808 L-210, Livermore, CA 94551, USA.\\
Institute for Computational Cosmology,
Department of Physics,
Durham University,
South Road, Durham DH1 3LE, UK.}
\email{schneider42@llnl.gov}
\author{Shaun Cole and Carlos S. Frenk}
\affil{Institute for Computational Cosmology,
Department of Physics,
Durham University,
South Road, Durham DH1 3LE, UK.}
\author{Istvan Szapudi}
\affil{Institute for Astronomy, University of Hawaii, 2680 Woodlawn Drive, Honolulu, HI 96822, USA.}

%\date{Received \today}

%\pagerange{\pageref{firstpage}--\pageref{lastpage}} \pubyear{2010}

%\label{firstpage}

\begin{abstract}
We present an algorithm for quickly generating multiple realizations of $N$-body simulations to 
be used, for example, for cosmological parameter estimation from surveys of large-scale structure.
Our algorithm uses a new method to resample the large-scale (Gaussian-distributed) Fourier modes in a 
periodic $N$-body simulation box in a manner that properly accounts for the nonlinear mode-coupling between 
large and small scales.
We find that our method for adding new large-scale mode realizations recovers the nonlinear power 
spectrum to sub-percent accuracy on scales larger than about half the Nyquist frequency of the simulation 
box.  Using 20 $N$-body simulations, we obtain a power spectrum covariance matrix estimate that matches 
the estimator in~\citet{takahashi09} (from 5000 simulations) with $<20$\% errors in all matrix elements.  
Comparing the rates of convergence, we determine that our algorithm requires $\sim 8$ times fewer simulations 
to achieve a given error tolerance in estimates of the power spectrum covariance matrix.
The degree of success of our algorithm indicates that we understand the main physical 
processes that give rise to the correlations in the matter power spectrum.  
Namely, the large-scale Fourier modes modulate both the degree of structure growth 
through the variation in the effective local matter density and also the spatial frequency of small-scale 
perturbations through large-scale displacements.
We expect our algorithm to be useful for noise modeling when constraining cosmological parameters from 
weak lensing (cosmic shear) and galaxy surveys, rescaling summary statistics of $N$-body simulations for 
new cosmological parameter values, and any 
applications where the influence of Fourier modes larger than the simulation size must be 
accounted for.  
\end{abstract}


\maketitle

%%%%%%%%%%%%%%%%%%%%%%%%%%%%%%%%%%%%%%%%%%%%%%%%%
\section{Introduction}
\label{sec:introduction}
The matter power spectrum is a sensitive probe of cosmological models.  
Observations of galaxy clustering, weak gravitational lensing, and the Lyman-$\alpha$ forest are all tracers of the matter power spectrum and have to-date provided key constraints on the initial conditions of the universe as well as on the growth and expansion history.  Many future surveys will rely on tracers of the matter power spectrum to constrain models for dark energy, dark matter, and massive neutrinos.  However, due to nonlinear gravitational evolution of the matter distribution, both the mean matter power spectrum as well as its variance and correlations between bands in wavenumber are difficult to predict without expensive numerical simulations.  

Considerable work has been done to predict the mean nonlinear matter power spectra to the precision necessary for near-term surveys~\citep[e.g.,][]{smith03, lawrence10}.  In order to use observations of the matter power spectrum to constrain cosmological parameters it is necessary to know the error distribution for the power spectrum estimates in addition to the mean spectrum predicted by the cosmological model.  It is by now well established that nonlinear gravitational evolution creates a complicated error distribution for the power spectrum with signficantly increased small-scale variance and correlations between all band powers when compared with the Gaussian case~\citep{scoccimarro99, meiksin99,hamilton06,neyrinck06,neyrinck07,takahashi09}.  The differences between the Gaussian error and nonlinear error models for the matter power spectrum are quite significant for parameter estimation from galaxy surveys~\citep{hamilton06, neyrinck07}, but uncertainty in the nonlinear galaxy bias with respect to the dark matter may dominate the inferred parameter constraints in the near future.  Weak lensing surveys probe the matter distribution directly and are thus insensitive to biasing.  The line-of-sight projection in weak lensing reduces the nonlinear contribution to the power spectrum covariance, but it is still a significant consideration for parameter estimation~\citep{semboloni07, takada09, sato09}.

Standard estimators of the matter power spectrum covariance matrix 
in a given model require a large ensemble of 
statistically independent realizations of the matter density over volumes at least as large as the
survey being analyzed.  \citet{meiksin99} found that at least several hundred realizations are 
necessary to estimate the covariance over an interesting dynamical range and~\citet{takahashi09} 
have used 5000 simulations to obtain a low-noise estimator for a Gpc$^{3}$ volume.
\citet{hamilton06} attempted to estimate the covariance matrix by sub-sampling a single 
simulation volume, but found that the window functions they applied to define different 
pseudo-independent subvolumes significantly altered the covariance.  That is, the Fourier transform 
of a windowed density field has Fourier modes that are linear combinations of the modes in the 
$N$-body simulation volume with periodic boundary conditions.  The modes of the windowed density
then have different covariance than the modes in the periodic box, which is particularly enhanced 
by the correlations between the largest and smallest scales~\citep{rimes06}. 
In a similar line of inquiry, \citet{norberg08} showed that jacknife estimates of the covariance 
matrix from simulated survey regions systematically underpredict the true nonlinear covariance 
matrix.  Finally, if some information is known about the structure of the power spectrum covariance,
then the ``shrinkage estimator'' of~\citet{pope08} gives a way to reduce both the bias and noise 
in covariance estimates from a fixed number of density realizations.  However, in our experience the 
``shrinkage estimators'' as formulated in \citet{pope08} rely quite heavily on the prior information 
one is able to supply.

In this paper we describe a method to obtain multiple realizations of the matter density from a 
single $N$-body simulation by resampling those Fourier modes (in a periodic box) that can be 
approximated as Gaussian distributed with zero correlations between Fourier modes.  While 
we anticipate this algorithm could be useful for a number of applications~\citep[such as adding 
power larger than the simulation volume as in][]{cole97}, in this paper we focus on the 
minimum number of simulations needed to estimate the matter power spectrum covariance matrix.
As a benchmark for our study we use the results of~\citet{takahashi09}, who used 5000 $N$-body 
simulations to estimate the covariance matrix in a 1~($h^{-1}$Gpc)$^{3}$ simulation volume.  
While we focus here on the 3-D matter power spectrum, one final goal will be to efficiently 
estimate the weak lensing power spectrum covariance for upcoming cosmic shear surveys.  
An example of this using a brute-force approach is given in~\citet{sato09}.  

The structure of this paper is as follows.  In Section~\ref{sec:resampling} we describe our method 
for resampling large-scale modes in a periodic $N$-body simulation and the necessary adjustments 
to the remaining small-scale modes.  
We give an approximate algorithm requiring only a single snapshot, but limited to simulations 
of at least several Gpc on a side, in Section~\ref{sec:singlesnapshotalgorithm}.  
We describe our full algorithm for arbitrary simulation sizes (as long as the large-scale modes 
are Gaussian distributed) in Section~\ref{sec:fullalgorithm}.
In Section~\ref{sec:validation} we present validation 
tests for adding new large-scale mode realizations.  
We describe how our resampling algorithm can be useful for 
power spectrum covariance matrix estimation in Section~\ref{sec:covestimation}.
We discuss several future directions and 
applications for this work in Section~\ref{sec:conclusions}.  
Where numerical results are given, unless otherwise stated, 
we assume the same cosmology as~\citet{takahashi09} with
$\Omega_{m}=0.238$, $\Omega_{b}=0.041$, $\Omega_{\Lambda}=0.762$, 
$n_{s}=0.958$, $\sigma_{8}=0.76$, and $H_{0} = 73.2~{\rm km}\,{\rm s}^{-1}\,{\rm Mpc}^{-1}$.

%%%%%%%%%%%%%%%%%%%%%%%%%%%%%%%%%%%%%%%%%%%%%%%%%
\section{Resampling large-scale power}
\label{sec:resampling}
Our goal is to isolate and resample those Fourier modes in a single periodic $N$-body simulation
that can be accurately approximated as Gaussian distributed.  
We assume that the Fourier modes of the initial conditions
for the simulation are generated from a random realization of a complex Gaussian field on a grid from a 
specified power spectrum ({\it i.e.} the real and imaginary parts of the Fourier modes are each drawn from 
independent Gaussian distributions such that the amplitude of the modes is equal to the square root of the 
power spectrum).  As the simulation is evolved, gravitational interactions between the particles tracing the 
matter density field cause the phases of the Fourier modes to become correlated and the distribution
of Fourier mode amplitudes to become skewed towards larger values.  The growth rate of correlations 
and departure from Gaussianity are functions of wavenumber as gravitational collapse proceeds
most rapidly on small-scales.

So, we first need to separate the Fourier modes into ``large-scales'' defined to be Gaussian distributed, and 
``small-scales'' that we will define to be all the remaining Fourier modes in the simulation.  Because it 
worked for our purposes we crudely separate large and small scale modes by defining 
all modes with wavenumber amplitude less than some value $\kthresh$ as Gaussian distributed.  
In practice we defined $\kthresh$ so that the r.m.s. mass overdensity smoothed in top-hat spheres 
of radius $2\pi/\kthresh$ was less than 0.2~\citep[similar to][]{cole97}.  This is based on the idea that larger
mass overdensity fluctuations indicate more nonlinear growth leading to larger departures from 
Gaussianity.

Having defined those Fourier modes that we will model as Gaussian, we next resample them 
by generating new realizations of the initial conditions 
that have nonzero Fourier modes only for $|\kv| < \kthresh$
and extrapolating the amplitudes of the modes to the desired redshift using linear theory.

To obtain a new realization of the density field that approximates what
we would have obtained had we run a simulation with the new large-scale modes in the 
initial conditions of the simulation, 
we have to adjust the small-scale modes in some way to account for the 
phase correlations that would have emerged between the large and small scales.

\subsection{Resampling algorithm}
\label{sec:modeaddition}
In this Section we imagine that we know the particle positions in a simulation 
that has no large-scale power (where ``large-scale'' 
denotes Fourier modes with $|\kv| \leq \kthresh$) and we seek an algorithm to add 
new large-scale modes.  We will denote the simulation without large-scale power 
as a ``small-modes'' simulation and the simulation with all Fourier modes nonzero as 
an ``all-modes'' simulation.

\subsubsection{Physical picture}
 Altering the amplitudes and phases of the large-scale modes of the mass density in a
cosmological volume has two dominant effects, 
\begin{enumerate}
\item sub-volumes are expanded or compressed,
\item sub-volumes gravitationally evolve with an effective
matter background density that is enhanced or diminished, 
\end{enumerate}
depending on where the large-scale modes become more or less overdense,
respectively~\citep{frenk88,little91,tormen96, cole97,angulo10}.  
We follow~\citet{cole97} to model both of these effects.  
To model the expansion and compression of 
sub-volumes, we perturb the particle positions using the
Zel'dovich displacements~\citep{zeldovich70} 
calculated from the large-scale modes, $\delta_L$,
\begin{equation}\label{eq:zeldovich}
  \dvec_{L}(\xv,a) \equiv \int \frac{d^3k}{(2\pi)^3}
  \frac{\delta_L(\kv,a)\kv}{k^2}\,
  e^{i\kv\cdot\xv}.
\end{equation}
The expansion and compression of sub-volumes is analogous to the
frequency modulation of a radio wave.  The large-scale modes to be 
added are a new signal that modulates the sizes of sub-volumes in 
the matter density field; expressed to first order by the 
Zel'dovich displacements in equation~(\ref{eq:zeldovich}).

\begin{figure}
  \centerline{
    \includegraphics[scale=0.48]{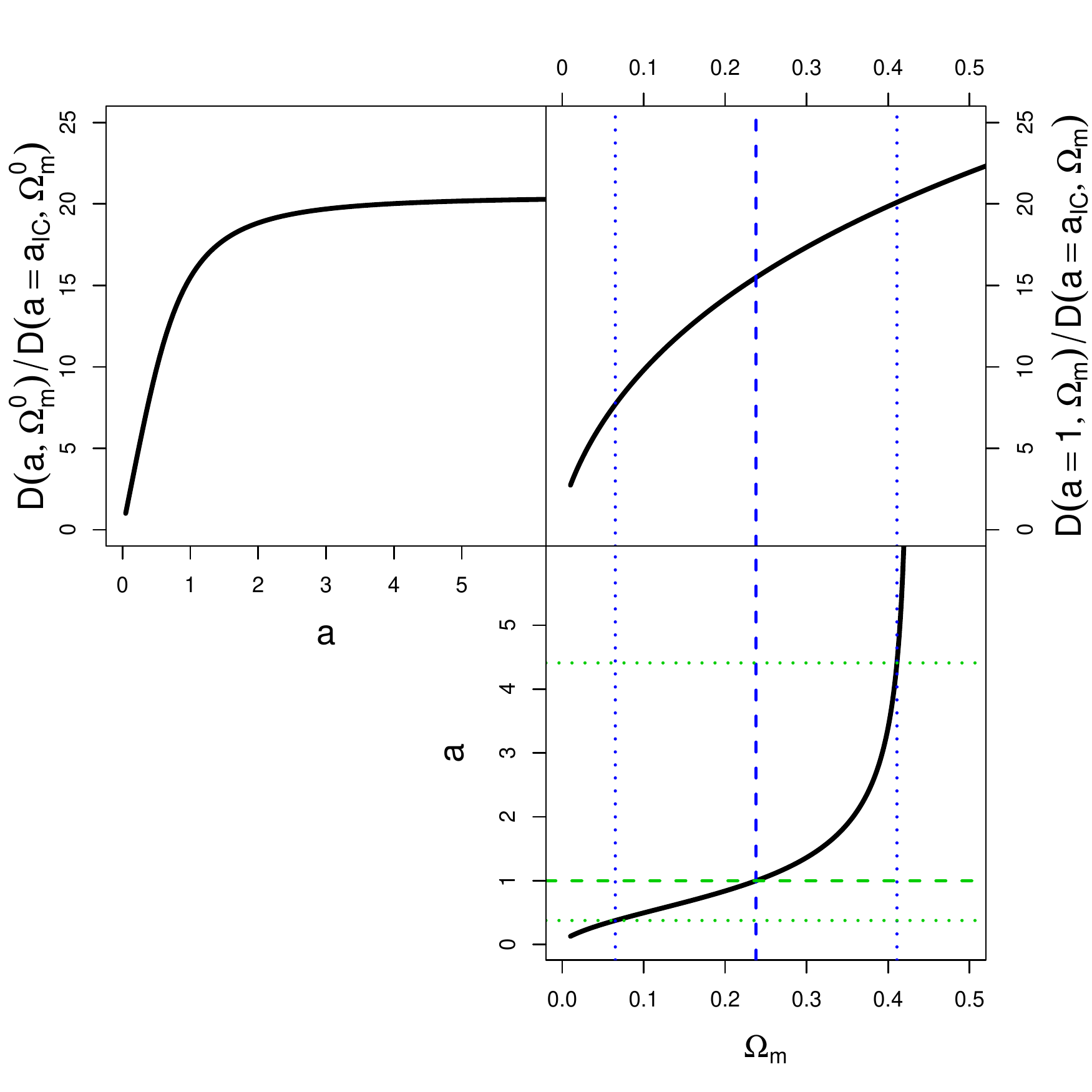}
  }
  \caption{\label{fg:scalefactor_vs_omegam}Top left: Linear growth function normalized 
  to one at the time of the simulation initial conditions ($a_{\rm IC}=1/21$) versus scale factor.
  Top right: Linear growth function versus matter density, also normalized to unity at the 
  time of the initial conditions for all values of $\Omega_{m}$.  Bottom: scale factor as a 
  function of matter density at matched linear growth.  The vertical dashed line shows the 
  value of $\Omega_{m}=\Omega_{m}^{0}=0.238$ used in our simulations.  
  The vertical dotted lines show the values 
  of  $\Omega_{m}^{0}(1\pm4.3\sigma(\delta_{L}))$, where $\sigma(\delta_{L})\approx0.17$ is the r.m.s. 
  mass over-density in top-hat spheres  at the scale $R=2\pi/(256 k_{F})$ ($k_F\equiv 2\pi/\lbox$) for 
  the density perturbations with nonzero Fourier amplitudes 
  only for $|\kv| < \kthresh$.  The horizontal dotted lines
  in the bottom panel show the corresponding scale factor range over which our small-modes 
  simulation must be saved.}
\end{figure}

Compressed (expanded) regions also evolve with an increased (decreased) 
effective matter density, altering the growth rate of structure.  
Because the growth rate increases monotonically with time, we 
model the changing effective local matter density 
by finding new times $a'(\xv)$ at which the linear growth rate matches 
the growth rate in a universe with matter density 
$\Omega_{m}\left(1+\delta_{L}(\xv, a)\right)$ (with $\Omega_{\Lambda}$ kept fixed). 
The perturbed scale factors $a'(\xv)$ are then defined by the relation,
\begin{equation}\label{eq:growthmatching}
  D(a',\Omega_m) \approx D(a, \Omega_m \left(1+\delta_L(\xv,a)\right) ),
\end{equation}
where $D(a,\Omega_{m})$ is the linear growth function. 
In practice, we solve equation~(\ref{eq:growthmatching}) numerically to get $a'$ 
as demonstrated in Figure~\ref{fg:scalefactor_vs_omegam}.

% We then move each particle in the simulation to its position at $a'$; 
% found by interpolating between simulation snapshots.

\subsubsection{Approximate method for a single snapshot}
\label{sec:singlesnapshotalgorithm}
Combining the two effects of expansion/compression and time-perturbations, 
the model for adding large-scale fluctuations to the mass density 
field is,
\begin{equation}\label{eq:densitymodel}
	\delta(\xv, a) = \delta_{L}(\xv,a) + \delta_{S}(\xv', a'),
\end{equation}
where $\delta(\xv,a)\equiv\rho(\xv,a)/\rhobar - 1$ is the mass over-density that would 
be measured in the all-modes simulation, $\delta_{S}(\xv,a)$ is the 
mass over-density measured in the small-modes simulation, $\delta_{L}$ has zero power for 
$|\kv| > \kthresh$, $a'$ is defined by equation~(\ref{eq:growthmatching}),
and $\xv' \equiv \xv + \dvec_{L}(\xv,a)$.
Note that for this definition of $\xv'$ the Zel'dovich displacements 
are added to the particle positions prior to application of the time-perturbation 
(i.e., at the positions at time $a$, not at $a'$).  
We will discuss this choice below.

When $|a' - a| \ll 1$ (which requires $\left< \left|\delta_L(\xv)\right|^{2}\right>^{1/2} \ll 1$) 
the first order correction to $\delta_{S}$ in $a'$ should be 
sufficient for adding $\delta_{L}$ to the density field.  
Expanding both sides of equation~(\ref{eq:growthmatching})
to first order in $(a'-a)$ and $\delta_L$,
\begin{align}
  %\delta_S(\xv,a') &= \delta_S(\xv,a) + (a'-a)\frac{d\delta_S(\xv,a)}{d a}
  %+ \cdots
  %\notag\\
  D(a') &\approx D(a) + (a'-a)\frac{dD}{da}
  \\
  D(a,\Omega_m(1+\delta_L(\xv,a))) &\approx D(a,\Omega_m) + 
  \Omega_m\delta_L(\xv,a)\frac{dD}{d\Omega_m}.\notag
\end{align}
Solving for $a'$ gives,
\begin{equation}\label{eq:timeperttaylor}
  \frac{a'}{a} -1\approx \delta_L(\xv,a) 
 \left(\frac{d\ln D}{d\ln\Omega_{m}} / \frac{d\ln D}{d\ln a}\right),
\end{equation}
which shows that the perturbed scale factor is roughly proportional to the 
large-scale density perturbations being added.
Using equation~(\ref{eq:timeperttaylor})
in the Taylor expansion of equation~(\ref{eq:densitymodel}) gives,
\begin{multline}\label{eq:deltataylor}
	\delta(\xv,a) \approx \delta_{L}(\xv,a) + \delta_{S}(\xv',a) + \\
	\delta_{L}(\xv,a)\, \frac{\partial\delta_{S}(\xv',a)}{\partial\ln a}\, 
	 \left(\frac{d\ln D}{d\ln\Omega} / \frac{d\ln D}{d\ln a}\right).
\end{multline}
This suggests that the addition of large-scale modes can be quickly 
computed given an estimate of the nonlinear growth rate of the small-modes simulation.
With a single snapshot, the growth rate can be estimated from the 
continuity equation\footnote{Note that for simulation volumes $\gg 1$~Gpc on a side
relativistic corrections to the continuity equation could be important.  
These arise from the way that 
the large scale potential perturbations modify the geodesic equations.},
\begin{equation}
    H(a)\frac{\partial\delta(\xv,a)}{\partial\ln a} = 
    -\frac{1}{a}\nabla\cdot
    \left((1+\delta)\mathbf{v}\right) ,
		%-3\dot{\Phi}.
\end{equation}
and, using equation~(\ref{eq:densitymodel}),
\begin{equation}
	\frac{\partial\delta_S(\xv,a)}{\partial\ln a} = 
  \frac{\partial\delta(\xv,a)}{\partial\ln a}
  - \frac{d\ln D}{d\ln a}\delta_L(\xv,a).
\end{equation}
So, given particle positions and velocities at scale factor $a$ for a simulation 
that has no large-scale modes, equation~(\ref{eq:deltataylor}) provides a method 
to add new large-scale mode realizations when the large-scale density perturbations 
are small.

According to equation~(\ref{eq:timeperttaylor}) the range of $a'-a$ is proportional to the 
range of $\delta_{L}(\xv, a)$.  Because $\delta_{L}(\xv,a)$ is Gaussian distributed 
by construction, we can determine the range of $\delta_{L}$ from its variance,
\begin{equation}\label{eq:sigmatophat}
	\sigma^{2}\left(\delta_{L}(\xv,a)\right) \equiv
	\int_{k_{F}}^{\kthresh} d\ln k\, \Delta^{2}(k)\, 
	W_{\rm top-hat}^{2}(k R),
\end{equation}
where $k_{F}\equiv 2\pi/\lbox$ is the fundamental frequency of the simulation box,
$\Delta^{2}(k)\equiv k^3 P(k) / (2\pi^2)$ is the theoretical dimensionless matter 
power spectrum, and $W_{\rm top-hat}(k R)$ is the Fourier transform of a spherical top-hat 
window of radius $R$.
The scale $R$ is roughly given by either the pixel scale of the gridded density field
or the mean inter-particle spacing, whichever is larger.  For the simulations we use 
below, $\lbox=1000$~\hmpc, $\kthresh=8 k_{F}\approx 0.05h^{-1}$Mpc, and, with 256$^{3}$ 
particles, $R\sim 2\pi/(256 k_{F})$.
Evaluating equation~(\ref{eq:sigmatophat}) gives $\sigma\approx 0.17$ which is 
still less than $0.2$ as we required when choosing $\kthresh$.  But, as shown in 
Figure~\ref{fg:scalefactor_vs_omegam}, the variations in $\delta_{L}(\xv,a)$ greater than 
about $2\sigma$ lead to $|a'-a| > 1$ making equation~(\ref{eq:deltataylor})
insufficiently accurate.  In our test simulations $\delta_{L}$ actually varies over 
$\pm 4.3\sigma \approx \pm0.87$, giving a range of perturbed scale 
factor $0.26 \lesssim a' \lesssim 4.4$ (shown by the dotted lines in
Figure~\ref{fg:scalefactor_vs_omegam}).

\begin{figure}
	\centerline{
		\includegraphics[scale=0.48]{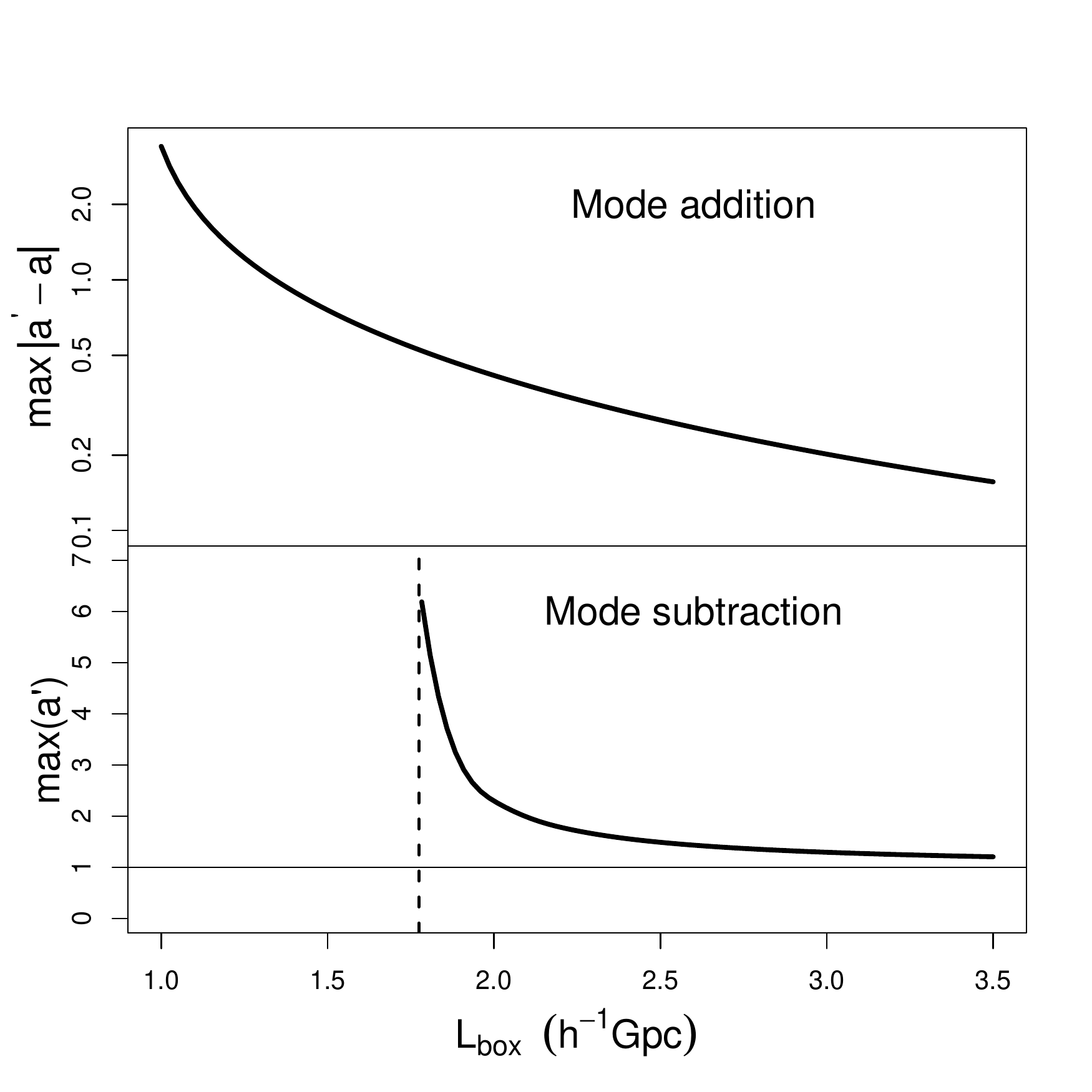}
	}
	\caption{Maximum variation in perturbed scale factor $a'$ as a function 
	of the simulation box size for fixed $\kthresh/k_{F}=8$ and assuming the maximum 
	fluctuations in $\delta_{L}(\xv,a)$ are $4.3\sigma$.  Top: Maximum absolute difference 
	between $a'$ and $a=1$ for the growth matching condition for adding large-scale 
	modes to a simulation (equation~\ref{eq:growthmatching}).  
	Bottom: Maximum $a'$ for the growth matching condition 
	for removing large-scale modes from a simulation (equation~\ref{eq:inversegrowthmatching}).
	There is no solution for $a'$ left of the vertical dashed line.}
	\label{fg:avslbox}
\end{figure}

\begin{figure}
	\centerline{
		\includegraphics[scale=0.48]{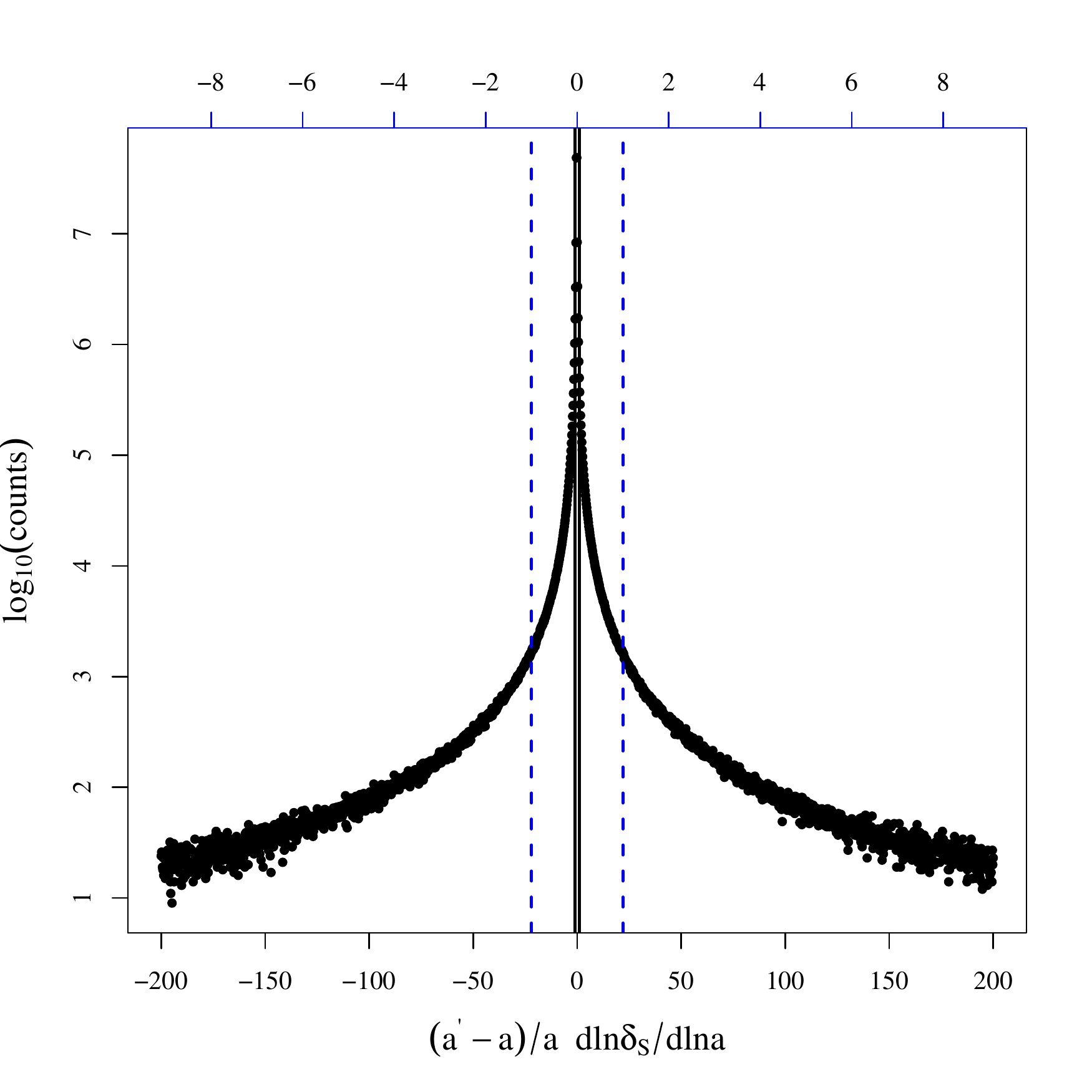}
	}
	\caption{Histogram of the values of $\frac{(a'-a)}{a}\,\frac{d\ln\delta_S}{d\ln a}$
	on a 512$^3$ grid for our $\Lambda$CDM cosmology in a 1~($h^{-1}$Gpc)$^3$ simulation.
	The tick marks on the top (blue) axis show the projected bin values if we 
	scale the bins on the bottom axis by the ratio of the maximum perturbed scale factor, 
	${\rm max}(a'-a)$,
	in a 3~$h^{-1}$Gpc box versus the maximum perturbed scale factor in our
	1~$h^{-1}$Gpc box.  The vertical dashed lines show the values of $\pm 1$ on this axis,
	where the third term in equation~(\ref{eq:deltataylor}) would be small relative to $\delta_S$.}
	\label{fg:nonlineargrowthhist}
\end{figure}

In Figure~\ref{fg:avslbox} we show the maximum value of $|a'-a|$ as a function 
of the simulation box size while keeping $\kthresh/k_{F}$ fixed at 8 and, as in the preceding
paragraph, assuming maximal variations in $\delta_{L}(\xv,a)$ of $4.3\sigma$.
For $\lbox \sim 3000$\hmpc, as in the recently completed MXXL 
simulation\footnote{The MXXL simulation (Angulo et al. in preparation) 
was run by the Virgo Consortium (\url{http://www.virgo.dur.ac.uk})
in the spring of 2011.  It is 3~$h^{-1}$Gpc on a side with $3\times 10^{11}$ particles 
with mass $6.3\times 10^9\,M_{\odot}$.},
the variation in $a'-a$ is less than 0.2 and the 
approximate formula in equation~(\ref{eq:deltataylor}) could be sufficient for adding the 
large-scale modes to the small-modes simulation.
To further assess the validity of equation~(\ref{eq:deltataylor}), we show a histogram of the 
values of the third term in equation~(\ref{eq:deltataylor}), normalized by $\delta_S(\xv',a)$,
on a 512$^3$ grid in our 1~$h^{-1}$Gpc small-modes simulation in Figure~\ref{fg:nonlineargrowthhist}.
The distribution of grid cell values is strongly peaked around zero, but has wide tails 
extending to values of several hundred.  This again illustrates that equation~(\ref{eq:deltataylor})
is a bad approximation for a (1~$h^{-1}$Gpc)$^3$ volume.  However, if we scale the 
bin values on the lower horizontal axis by the ratio of the maximum perturbed scale 
factor ${\rm max}(a'-a)$ in a (3~$h^{-1}$Gpc)$^3$ simulation to the maximum $(a'-a)$ in 
our simulation, we get the axis on the top of Figure~\ref{fg:nonlineargrowthhist}.  
For a (3~$h^{-1}$Gpc)$^3$ simulation 92\% of the grid cells of the third term in 
equation~(\ref{eq:deltataylor}) have values less than one.  So, equation~(\ref{eq:deltataylor}) may 
be a good enough approximation for adding large-scale modes to the MXXL or other 
simulations with similarly large volumes.

\subsubsection{Full recipe}
\label{sec:fullalgorithm}
For box sizes smaller than several Gpc on a side, we need an algorithm that 
accurately implements the growth matching condition of equation~(\ref{eq:growthmatching})
to compute the density $\delta_S(\xv',a')$ in equation~(\ref{eq:densitymodel}).
For computational expediency and for later comparison 
with~\citet{takahashi09}, we have limited our test simulations to 
$\lbox=1000$\hmpc.  
Therefore, we have not tested the 
accuracy of equation~(\ref{eq:deltataylor}).  
Instead, because our test simulation 
parameters yield $|a'-a|>1$, we apply the model of equation~(\ref{eq:densitymodel})
by adjusting the positions of each particle in the simulation.
If $\xv_{i}(a)$ is the position of the $i$th particle at scale-factor $a$ in 
the small-modes simulation, the model from equation~(\ref{eq:densitymodel}) 
with modes $\delta_{L}$ added to the simulation
can be rewritten for each particle position as,
\begin{equation}
	\xv_{i}'\left(a'; \delta_{L}\right) \equiv \left[ \xv_{i}(a) + \dvec_{L}\left(\xv_{i}(a), a^{*}\right)\right]_{a = a'},
\end{equation}
where $a^*$ is the scale factor of the
target time where the full
density field (with new large-scale modes) is desired.  
That is, we first apply the same Zel'dovich displacement to the particle positions in all 
the snapshots (evaluated at the unperturbed positions in each snapshot).
Then, we interpolate between these modified snapshots to find the particle position
at perturbed time $a_{i}'  \equiv a'(\xv_{i} + \dvec(\delta_{L}(\xv_{i}(a)), a^{*}))$.
To find the particle positions at all values of $a'$ we must save simulation 
snapshots at times covering the range of $a'$ values and spaced sufficiently 
closely in $a$ so that the interpolation of positions is accurate.  This requires 
running the simulation far into the future ($a > 4$ for our test simulations) and 
saving $\sim 10$'s of snapshots, which creates both an extra computational 
and also a disk storage burden over the approximation in equation~(\ref{eq:deltataylor}).

The steps of our full algorithm for adding large-scale modes to the small-scale
density are then:
\begin{enumerate}
  \item Choose a threshold scale $\kthresh$ defining the boundary between large and small scale modes and a target time $a^*$ where the resampled density is desired.
  \item Calculate the range of perturbed scale factor $a'$ defined by
   equation~(\ref{eq:growthmatching}) about $a^*$ where the simulation snapshots will be required.
  \item Run an $N$-body simulation in a periodic volume with nonzero Fourier modes 
  in the initial conditions only for $|\kv| > \kthresh$ and save snapshots over the required 
  range of $a'$ that are spaced sufficiently close together to allow for accurate 
  interpolation of particle positions.
  \item Calculate a new realization of large-scale modes $\delta_L(a^*)$ on 
  a grid.
  \item Calculate the Zel'dovich displacements defined in equation~(\ref{eq:zeldovich}) from 
  $\delta_L$.
  \item Apply the Zel'dovich displacements to each snapshot of the small-modes 
  $N$-body simulation.
  \item Move each particle in the small=modes simulation to its location at 
  time $a'(\xv(a)+\dvec_L)$ as defined by equation~\ref{eq:growthmatching}, 
	interpolating between snapshots when necessary.
\end{enumerate}
The resulting particle positions give a close approximation to the particle positions 
that would be obtained by running an $N$-body simulation with the Fourier modes 
in $\delta_L$ present in the initial conditions (with the amplitudes suitably scaled 
according to the linear growth function).

The two operations 
of adding Zel'dovich displacements and perturbing
the time where the particle positions are evaluated do not commute
when the density field has evolved nonlinearly from the initial
conditions.  To decide on how to order these operations we imagine
an approximation to an $N$-body simulation where the initial conditions
are first evolved by adding Zel'dovich displacements to the particles 
and then the density is scaled by the linear growth function.
Reversing these operations, with the Zel'dovich move applied last,
would be less analogous to the way the particle positions 
actually evolve in a simulation. In this latter scenario, the 
particles would first gravitationally collapse at rates determined by the local 
mean matter density and then at late times would undergo displacements to 
expand or compress regions in over- or overdense volumes.  There is 
nothing physical about this ordering of operations.
We have confirmed with our test simulations
that the mode resampling does not work well unless the Zel'dovich move
is the first step of the mode-resampling algorithm.
%From this reasoning it might seem best to evaluate the Zel'dovich moves 
%at the time of the initial conditions rather than at the final time $a^{*}$.  
%But we have also verified with our test simulations that this does 
%not work very well.  
%Perhaps this could be because we have not constructed a model for the 
%perturbation of the density field that accounts for coupling between the 
%effects of expansion/compression and time-perturbation, but we will see 
%that the algorithm is suitable for our purposes without adding 
%further complications.

%\begin{align}
%  \left.\frac{d\,\delta_S\left(\xv'(a'),a'\right)}{d\ln
%      a'}\right|_{a'=a^*} &= 
%  \frac{\partial\,\delta_S(\xv'(a^*),a^*)}{\partial a} + 
%  \notag\\
%  &\qquad+\frac{\partial\,\delta_S(\xv'(a^*),a^*)}{\partial\,\xv}
%  \cdot \frac{\vv(a^*)}{H(a^*)}
%\end{align}

\subsection{Algorithm implementation}
\label{sec:implementation}
The algorithm in Section~\ref{sec:modeaddition} describes how to add large-scale 
Fourier modes to a simulation that had zero large-scale power in the 
initial conditions.  Ideally we would like to first remove the existing 
large scale modes from a previously run simulation and then apply
the algorithm in Section~\ref{sec:resampling}.  We will discuss why this
is challenging in Section~\ref{sec:subtractingpower}, 
but first we describe details of the implementation
of the mode addition algorithm.

To generate a simulation without large-scale power (what we call 
a ``small-modes'' simulation), we generated 
Gaussian Fourier mode realizations on a grid only for those modes, set all modes 
with $|\kv| < \kthresh$ to zero, and then performed the reverse 
Fourier transform to get the initial density field that is used for the 
first particle displacement step.  We then evolve the simulation normally, 
but saving 43 snapshots equally spaced in $\ln(a)$ over the range 
$0.1 \lesssim a \lesssim 5.4$ so the 
particle positions at perturbed times can be accurately determined 
by interpolating between snapshots.  
For testing purposes we ran another simulation using the 
same initial conditions, but before the large-scale modes were set to 
zero (called an ``all-modes'' simulation).  
The particle positions at $a=1$ in the all-modes simulation
give our benchmark for testing the reconstruction of the density field 
when the large-scale modes that were deleted from the initial conditions 
are added back into the small-modes simulation at $a=1$.
Note that all the phases in the initial conditions were identical for both 
the all-modes and small-modes simulations.
Unless otherwise specified, all our simulations were in box 
$1\, h^{-1}{\rm Gpc}$ on a side with $256^{3}$ particles and were 
evolved using a lean version of the Gadget-2 code~\citep{gadget2b,gadget2a} from Gaussian 
initial conditions at redshfit 20. 

\begin{figure}
	\centerline{
	\includegraphics[scale=0.5]{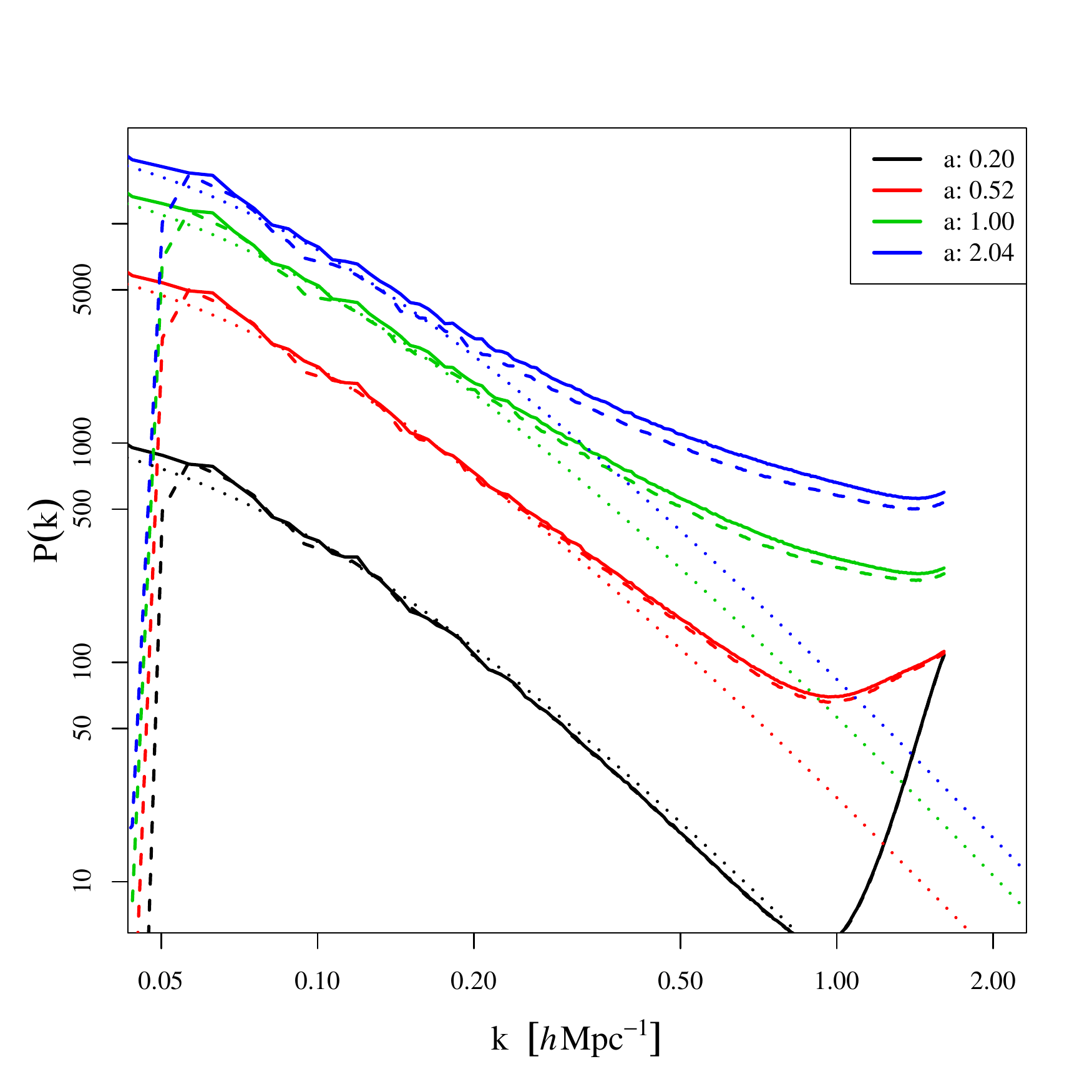}
	}
	\caption{\label{fg:psgrowthrates}Matter power spectra for several snapshots 
	of the simulations with scale factor increasing from bottom to top.  The solid lines
	are for the all-modes simulation, dashed lines are for the small-modes 
	simulation and dotted lines show the linear theory prediction.  Notice the 
	divergence of the nonlinear all-modes and small-modes spectra as 
	the simulation proceeds. }
\end{figure}

Because gravitational evolution couples small-scale Fourier modes to the 
largest modes in the simulation box, the nonlinear perturbation growth rate 
in the small-modes simulation (i.e., with large-scale modes missing) 
is smaller than that in the all-modes simulation, 
as illustrated with the power spectrum comparison in Figure~\ref{fg:psgrowthrates}.
The solid lines show the power spectra of select snapshots in the all-modes simulation
while the dashed lines show power spectra of the small-modes simulation at the 
same times (recall that the power spectra are equal in the initial conditions).
At late times, the all-modes power spectrum has a significantly larger amplitude 
on small scales than the small-modes power spectrum.  A first test of our 
mode-addition algorithm will be to correct for this deficit of nonlinear power.

\begin{figure}
  \centerline{
    \includegraphics[scale=0.45]{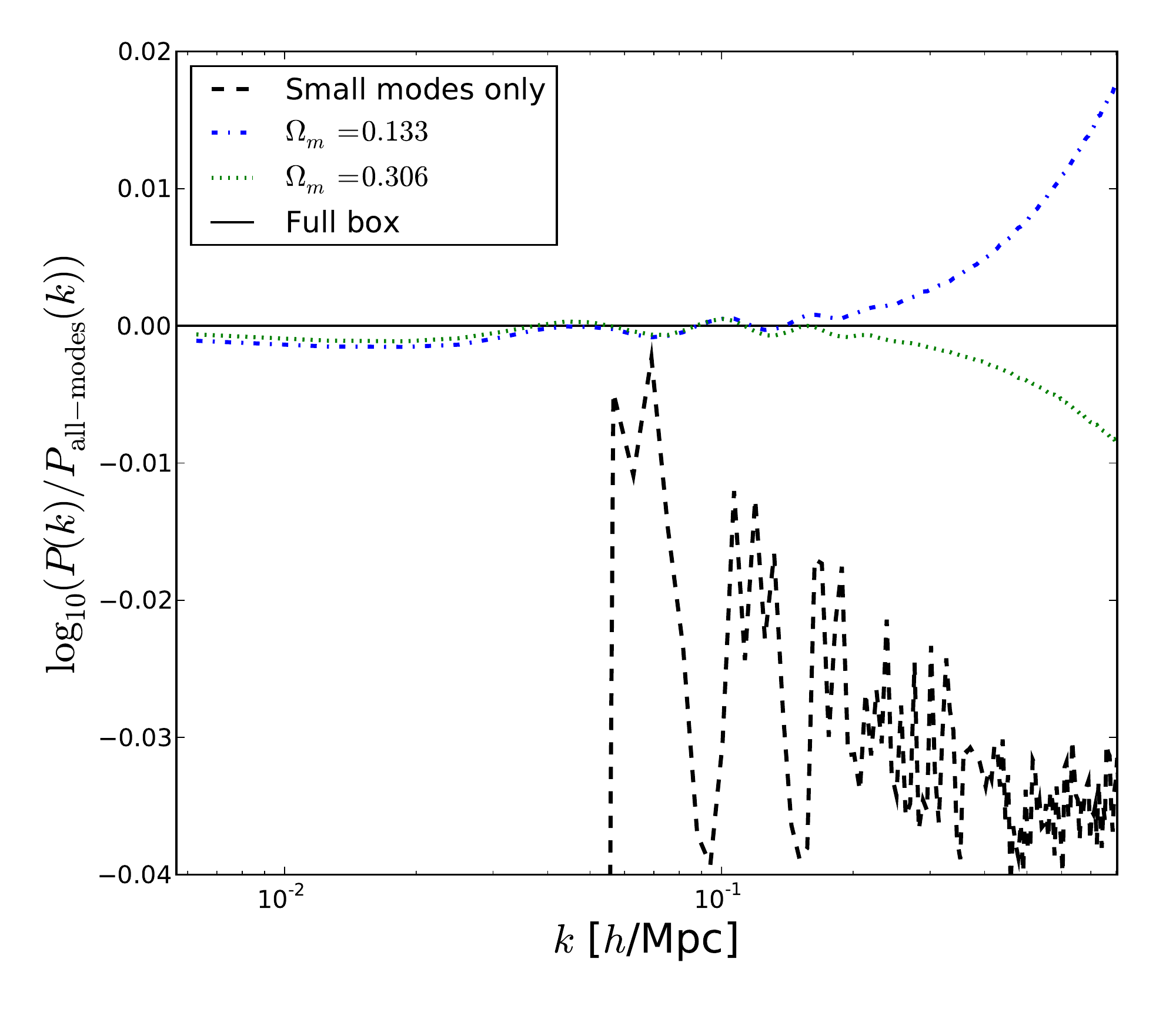}
  }
  \caption{\label{fg:omegam_match}Power spectra relative to the all-modes spectra 
	from simulations with
    different values of $\Omega_m$ plotted at matched values of the
    linear growth.  For comparison, the power spectra for our fiducial
  cosmology with and without large-scale modes are also shown.  The
  variation in the spectra with different $\Omega_m$ is only apparent
  for $k \gtrsim 0.5$~$h$Mpc$^{-1}$ and even there it is much less
  than the difference between the spectra with large-scale modes removed.}
\end{figure}

In order for the growth-matching condition in equation~(\ref{eq:growthmatching}) to 
provide a useful means for correcting for missing large-scale power, 
the errors in matching the growth rate by perturbing the matter density 
must be much smaller than the change in the growth rate due to the ommission 
of large-scale power.  We tested this condition explicitly by running
 two simulations with different values of
$\Omega_m$ and we plot their power spectra at matched linear growth in
Figure~\ref{fg:omegam_match}.  The power spectra in the two cosmologies 
match to $<0.1$\% for wavenumbers $k < 0.2$~$h/$Mpc, which is much 
less variation than between the all-modes and small-modes power spectra.  
This is why matching the linear 
growth is a good way to mimic the change in the local matter density.
We note that \citet{zheng02} have performed similar tests and showed that when the 
linear theory power spectrum is fixed (as we have assumed), 
several clustering statistics besides just the nonlinear power spectrum can 
be easily related between models with different $\Omega_m$ and $\sigma_8$.

\begin{figure*}
  \centerline{
    \includegraphics[scale=0.5]{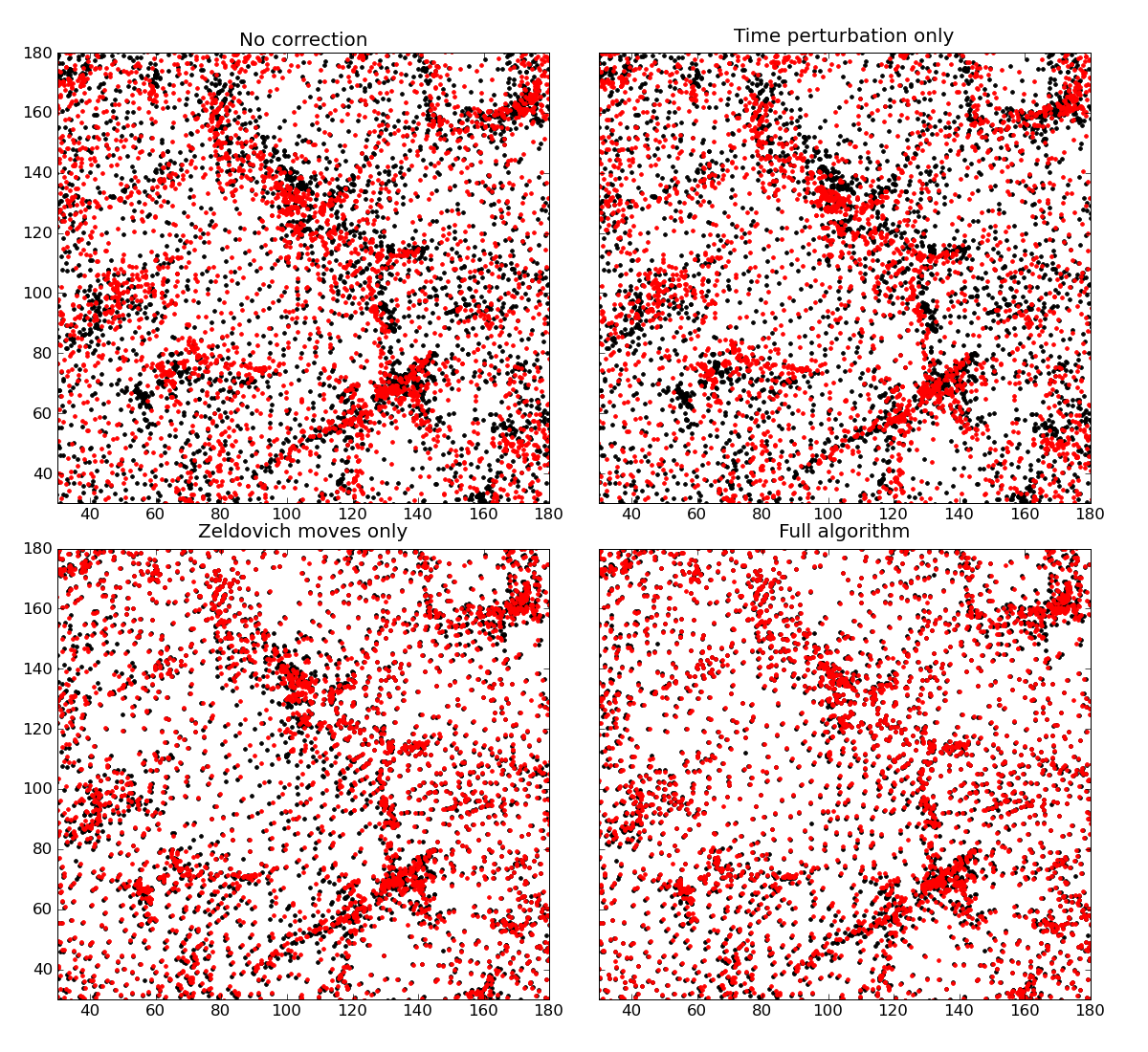}
  }
  \caption{\label{fg:dotplots} Particle positions in a section of our 1~Gpc$^{3}/h^{3}$ simulation box (of width 50~Mpc$/h$).  The red points denote the particle positions in a simulation with the large--scale Fourier amplitudes omitted from the initial conditions as described in the text.  The black points denote positions in a simulation with all Fourier modes included, but that is otherwise identical to the simulation denoted by the red points.  The 4 panels show various components of our algorithm for replacing the missing large--scale power.  The top left panel has no correction applied to the red points, the top right has the red particles shifted to their positions at different times so that the linear growth rate mimics the perturbation to the local matter density from the added large--scale modes.  The bottom left panel shows a Zel'dovich move applied to the red points, where the Zel'dovich displacements are derived from the large--scale modes only.  Finally the bottom right panel shows our full algorithm combining the time-perturbation and Zel'dovich displacements. 
  The axis labels are in $h^{-1}$Mpc.}
\end{figure*}

Next, we tested the various steps of the mode addition algorithm 
from Section~\ref{sec:modeaddition} by matching the particle positions between 
the all-modes and the small-modes simulations 
by adding the large-scale modes from the all-modes initial conditions 
back into the small-modes simulation at $a=1$ (after scaling the large-scale 
modes by the linear growth function).
Figure~\ref{fg:dotplots} illustrates our comparison for a section of the 
$1 (h^{-1}{\rm Gpc})^{3}$ simulation volume.  The black points show the 
particle positions in the all-modes simulation while red shows the positions 
in the small-modes simulation.
It is readily apparent to the eye that both the time-perturbation 
and Zel'dovich moves are necessary to obtain a good match to the particle 
positions in the all-modes simulation.
The Zel'dovich move alone does a better job matching the particle positions in the largest over-densities than 
the time-perturbation alone, but leaves significant position offsets compared to the final algorithm combining 
both perturbations.  

\subsection{Subtracting large-scale power}
\label{sec:subtractingpower}
\begin{figure}
  \centerline{
    \includegraphics[scale=0.45]{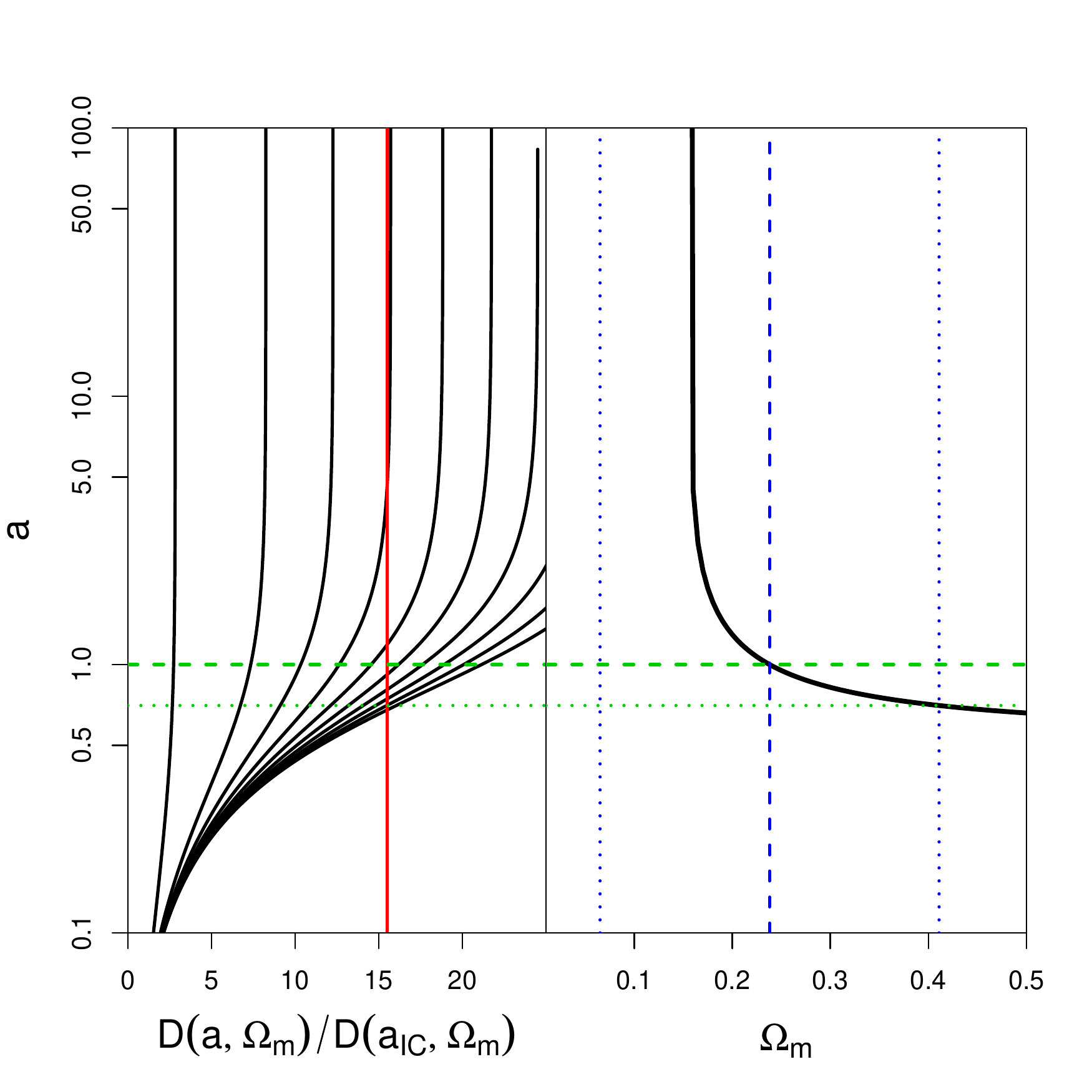}
  }
  \caption{\label{fg:growthtableinverse}Scale factor versus matter density at matched linear growth for mode subtraction.  Left panel: The scale factor versus the linear growth function normalized to one at the time of the initial
conditions ($a_{IC}=1/21$) for different values of $\Omega_{m}$ over the range $[0.01,0.5]$.  Each 
black line represents a different value of $\Omega_{m}$ starting with $\Omega_{m}=0.01$ for the left-most 
line and increasing in steps of 0.05.
The solid vertical red line shows the value of the normalized growth function at $a=1$ and
the target mass density $\Omega_{m}^{0}=0.238$.  
Right panel: The growth matching condition for removing large-scale modes in equation~(\ref{eq:inversegrowthmatching}) 
amounts to finding the value of the scale factor in the left panel where the red vertical line intersects the black lines 
for each $\Omega_{m}$ value.  The line of intersection points is plotted as the black line in the right panel.
The vertical dotted lines show the range of $\Omega_{m}^{0}(1\pm4.3\sigma(\delta_{L}))$ (as in 
Figure~\ref{fg:scalefactor_vs_omegam}).  The scale factor value matching the upper bound in 
the perturbation to $\Omega_{m}$ is shown by the dotted horizontal line.  There is no
scale factor value that can match the linear growth at the lower bound of the perturbed $\Omega_{m}$, 
rendering the mode subtraction impossible for the overdense regions of $\delta_{L}$.}
\end{figure}

The algorithm we have demonstrated requires that an $N$-body simulation be run with 
initial conditions that have the large-scale Fourier mode amplitudes set to zero and 
with a relatively large number of late time snapshots saved that can be used for 
interpolating particle positions.  Ideally, we would like to have an algorithm that 
can be applied to existing $N$-body simulations.  Such an algorithm would first 
require a method for subtracting the large-scale modes already present in the 
simulation.  Our procedure for adding large-scale modes should work in reverse, 
but with the new growth-matching condition, 
\begin{equation}\label{eq:inversegrowthmatching}
  D(a',\Omega_m (1 + \delta_{L}(\xv, a)) = D(a,\Omega_m).
\end{equation}
We arrive at this condition by modeling regions where $\delta_{L}$ is overdense (overdense) 
as regions that have evolved without large-scale power but with a larger (smaller) local value 
of $\Omega_{m}$.  
We then find the time 
$a'$ where a simulation with density $\Omega_m(1+\delta_{L}(\xv,a))$ and no large-scale 
power would have the same linear growth as the original simulation with density 
$\Omega_m$ at time $a$.  
However, removing the effect of large-scale under-densities can require evolving those overdense 
regions far into the future to match the growth on the right-hand side of equation~(\ref{eq:inversegrowthmatching}).
For $\Lambda$CDM, the local linear growth in large under-densities can freeze and will never 
match the right-hand side in equation~(\ref{eq:inversegrowthmatching}) even if evolved infinitely into the future.
The growth matching condition in equation~(\ref{eq:inversegrowthmatching}) and the problem 
with under-densities is illustrated in Figure~\ref{fg:growthtableinverse}.
% ----------
% ----------
For our fiducial cosmology, $\lbox=1\,h^{-1}{\rm Gpc}$, and $\kthresh=8 k_{F}$ 
all regions in $\delta_{L}$  that have density smaller than the mean density minus $\sim2\sigma$
will be impossible to correct using the growth matching condition of equation~(\ref{eq:inversegrowthmatching}).
In the bottom panel of Figure~\ref{fg:avslbox} we show the maximum $a'$  derived from 
equation~(\ref{eq:inversegrowthmatching}) as a function of simulation box size assuming 
$\kthresh / k_{F}=8$ and that the maximum variations in $\delta_{L}(\xv,a)$ are $4.3\sigma$.  
For cubic simulation boxes with sides less than $\sim 1.8~h^{-1}{\rm Gpc}$ there is 
no solution for $a'$.  For larger box sizes the maximum $a'$ quickly decreases from $\sim6$ to 
very close to one.  So for simulation box sizes of several $h^{-1}$Gpc, 
removing large-scale modes may be feasible with minimal extra computational effort.
However, again for computational expediency, we have limited our considerations in this 
paper to our $1~h^{-1}$Gpc simulations and therefore do not consider the 
removal of large-scale modes further.

%%%%%%%%%%%%%%%%%%%%%%%%%%%%%%%%%%%%%%%%%%%%%

%%%%%%%%%%%%%%%%%%%%%%%%%%%%%%%%%%%%%%%%%%%%%
\section{Validation tests}
\label{sec:validation}

We present further tests in this section to assess the accuracy of our 
method for introducing 
large-scale modes in $N$-body simulations.  We are primarily interested 
in the matter power spectrum and power spectrum covariance matrix 
and investigate estimators for these in turn.

\subsection{Recovering the nonlinear power spectrum}
\begin{figure*}
  \centerline{
    \includegraphics[scale=0.53]{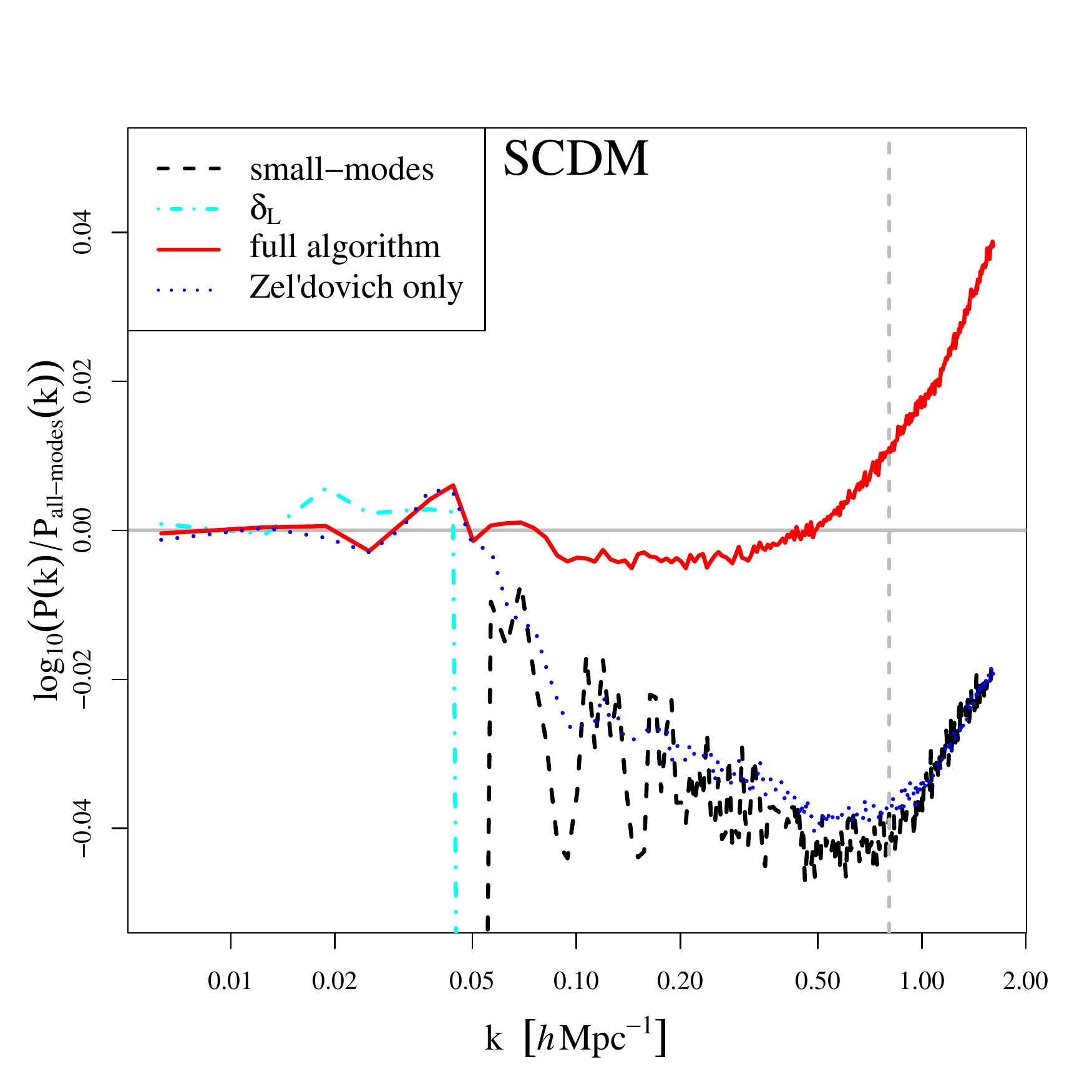}
    \includegraphics[scale=0.53]{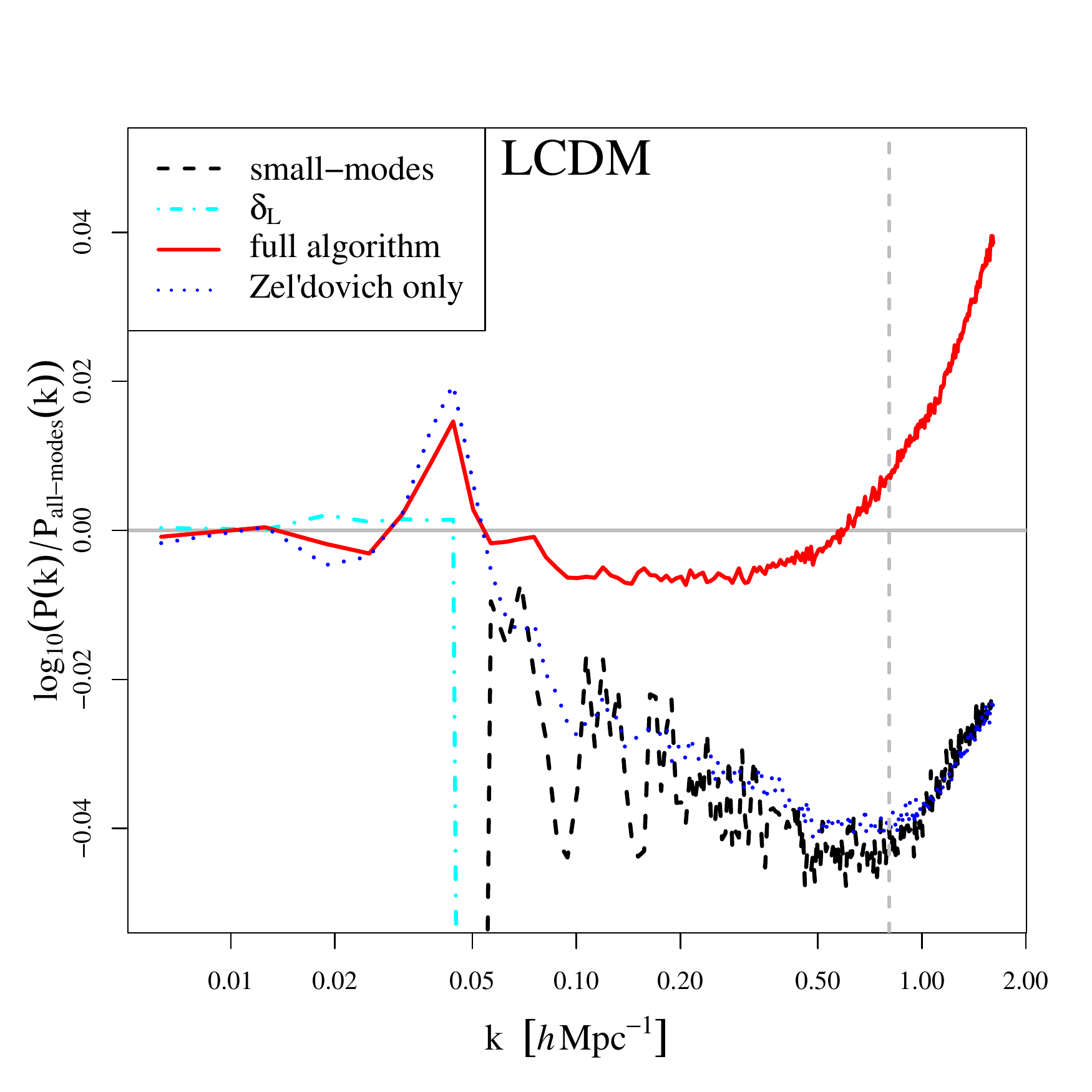}
  }
  \caption{\label{fg:addpowersnapshots}Power spectra relative to the all-modes spectra 
	demonstrating the
  accuracy with which large-scale power can be added to a simulation
  without any large-scale modes.  Left: SCDM, right: $\Lambda$CDM.  
  The large-scale modes ($\delta_{L}$) added to the simulations without 
  large-scale power (small-modes) have the same phases as in the simulation 
  run with all modes present (shown by the gray horizontal lines).  The vertical 
  dashed line marks the Nyquist frequency of the density fields.}
\end{figure*}

In Figure~\ref{fg:addpowersnapshots} we demonstrate the reconstruction of the 
nonlinear power spectrum from the small-modes simulation.  
Using the all-modes simulation as our benchmark, we used the 
Fourier modes from the all-modes initial conditions with $|\kv| < \kthresh$
to determine the Zel'dovich moves and perturbed times for adjusting 
the small-modes simulation (so if our algorithm worked perfectly the 
power spectra should be perfectly matched).
In addition to the $\Lambda$CDM simulations described in 
Section~\ref{sec:implementation}, 
we ran analogous all-modes and small-modes simulations for an 
SCDM cosmology (i.e., with $\Omega_{m}=1$).   
Because the SCDM model has no dark energy, the linear growth 
rate is completely determined by the matter density, and 
our growth matching condition in equation~(\ref{eq:growthmatching}) 
should be an especially good approximation for the effect of 
large-scale Fourier modes on the gravitational growth rate.
Figure~\ref{fg:addpowersnapshots} shows that the fractional errors in the reconstructed 
power spectra are indeed slightly smaller for the SCDM simulation, but the errors are much 
less than 1\% for both cosmologies.  Due in part to the cloud-in-cell deconvolution
(which over-corrects the power on small scales), 
the fractional errors begin to increase significantly at about $0.5\,k_{\rm Nyquist}$.
%The small deviations of the yellow dashed lines from zero in Figure~\ref{fg:addpowersnapshots} 
%indicate deviations from linear theory in the all-modes power spectra 
%at $a=1$ (which are used to normalize the yellow lines).  

Note that the Zel'dovich correction alone, 
shown by dotted blue lines in Figure~\ref{fg:addpowersnapshots},
does a very poor job of correcting the small-scale power spectrum 
for the added large-scale modes.  This seems to be in contrast to the 
lower left panel of Figure~\ref{fg:dotplots} where the Zel'dovich correction 
alone appears to correct the particle positions rather well.  We now see that 
the eye notices the agreement of large-scale structures more easily than the 
disagreement of small-scale structures in the dot plots of 
Figure~\ref{fg:dotplots}.  

\subsection{Power spectrum covariance matrix}
The gravitational growth of perturbations increases the small-scale 
power spectrum variance and induces significant correlations between 
power spectrum bands~\citep{scoccimarro99, meiksin99}.  
In this Section we test how well our mode 
addition algorithm can reproduce these effects.  

By perturbing a single small-modes simulation with independent 
Gaussian realizations of the large-scale modes, we can 
generate many realizations of the density field.  
These are not independent realizations because the small-scale modes are 
initially the same for each resampling.  
% But it was not obvious to us 
% at the outset of our calculations whether the coupling to large-scale fluctuations is 
% significant enough to generate pseudo-independent realizations for the purpose 
% of estimating the power spectrum covariance matrix.  
We use the covariance matrix estimated in~\citet{takahashi09} from 5000 $N$-body 
simulations to test our sample covariance estimates from simulations 
with resampled large-scale modes.  

\subsubsection{Sample covariance matrix estimator}
First we define the estimator for the covariance matrix that we use 
for our comparisons.

Let $\dk$ denote the amplitude at grid point $\kv_i$ of the Discret
Fourier Transform (DFT) of the density field from an $N$-body simulation
measured on a 3D uniform periodic grid. 
Then, the standard estimator for the power spectrum of the mode
amplitudes is obtained by averaging $\left|\dk\right|^2$ over a
spherical shell of constant radius $\left|\kv_i\right|$, 
\begin{equation}
  \phat_q =  \frac{V}{\Nkq} \sum_{\kv_i, \left|\kv_i\right|=k_q}  \left|\dk\right|^2,
\end{equation}
where $\Nki$ is the number of modes that fit in the shell.  
We follow~\citet{takahashi09} and set this shell thickness to  $0.01~h{\rm Mpc}^{-1}$.
If we generate $N_r$ realizations of the density field (e.g. by
running $N_r$ identical $N$-body simulations except with different
random number seeds to generate the initial conditions), the mean
power spectrum estimator for all the realizations is: 
\begin{equation}\label{eq:meanpowerspectrumestimator}
  \pbar_q = \frac{1}{N_r}\sum_{a=1}^{N_r}\,\phat_q^{a}
\end{equation}
The estimator for the covariance of the power spectrum from all the realizations is then:
\begin{equation}
  \hat{C}_{ij} = \frac{1}{N_r}\sum_{a=1}^{N_r}
  \left( \phat_i - \pbar_i \right)\,
  \left( \phat_j - \pbar_j \right).
\end{equation}
  
\citet{meiksin99}, \citet{norberg08}, and \citet{takahashi09} found that 
at least several hundred 
realizations of the matter density field are required to accurately 
estimate the sample covariance matrix.  
This means that hundreds (or thousands in some cases) of $N$-body 
simulations must be run with different random number seeds in order 
to obtain a power spectrum covariance matrix estimate that is sufficiently accurate
for cosmological parameter estimation.  

\subsubsection{Multiple power spectra from a single simulation box}
Because running hundreds or thousands of $N$-body simulations is 
often prohibitively expensive, 
previous studies have attempted to estimate the power spectrum
covariance from a single $N$-body simulation by applying window
functions to the gridded density field to sub-sample the simulation volume.  
\citet{hamilton06} found that because the windows are convolved with 
the Fourier modes of the periodic simulation box, the resulting 
power spectrum covariance matrix estimates are significantly biased 
with respect to the result obtained from an ensemble of periodic simulations.
Furthermore this bias cannot be simply accounted for by deconvolving the 
window functions from the power spectrum estimates, but is a result of 
higher order connected correlations entering the covariance matrix estimator, 
which \citet{hamilton06} call ``beat-coupling.''

Although direct sub-sampling of the simulation volume has been shown not 
to reproduce the covariance from an ensemble of simulations, the idea that different
regions of a large simulation volume provide statistically independent information about the 
small-scale covariance is sound if we accept the ergodic hypothesis and assume our 
simulation is big enough.  
%As such, we have (unsuccessfully) investigated related 
%methods for estimating the small-scale power spectrum covariance from a single simulation 
%volume.
To avoid the convolution in Fourier space imposed by the configuration-space 
windows of~\citet{hamilton06}, we attempted to sub-sample the Fourier modes that 
enter the shell-averaged power spectrum estimator in equation~(\ref{eq:meanpowerspectrumestimator}).
For shells with large wavenumbers there are thousands of modes used to estimate
the power spectrum.  The nonlinear power spectrum covariance depends on the connected 
four-point correlations of the Fourier modes, but with thousands of modes on a uniform grid 
there must be many redundant four-point configurations that could yield independent 
samples of the (shell-averaged) power spectrum covariance (after appropriate scaling 
by the number of modes used to estimate the mean power spectrum).  

The enumeration and counting of the four-point configurations in each shell is complicated, 
so we instead divided the surface of each wavenumber shell 
into uniform area HEALPix\citep{healpix} pixels
 and then selected only one wavevector within each pixel to obtain 
a power spectrum estimate.  For a given number of pixels covering the spherical shell, the 
minimum multiplicity of wavevectors in a pixel determines the number of pseudo-independent 
power spectrum estimates we can derive with this method.  
That is, if all pixels have at least $n$ wavevectors within their boundaries, 
then we can generate at most 
$n$ pseudo-independent realizations of the power spectrum estimates.  
We defined smaller (and therefore more) pixels for shells of increasing wavevector magnitude.
This ensured that the multiplicity of each pixel was greater than one for a wide range 
in wavevector magnitudes.

\begin{figure}%[htbp]
	\centerline{
	\includegraphics[scale=0.3]{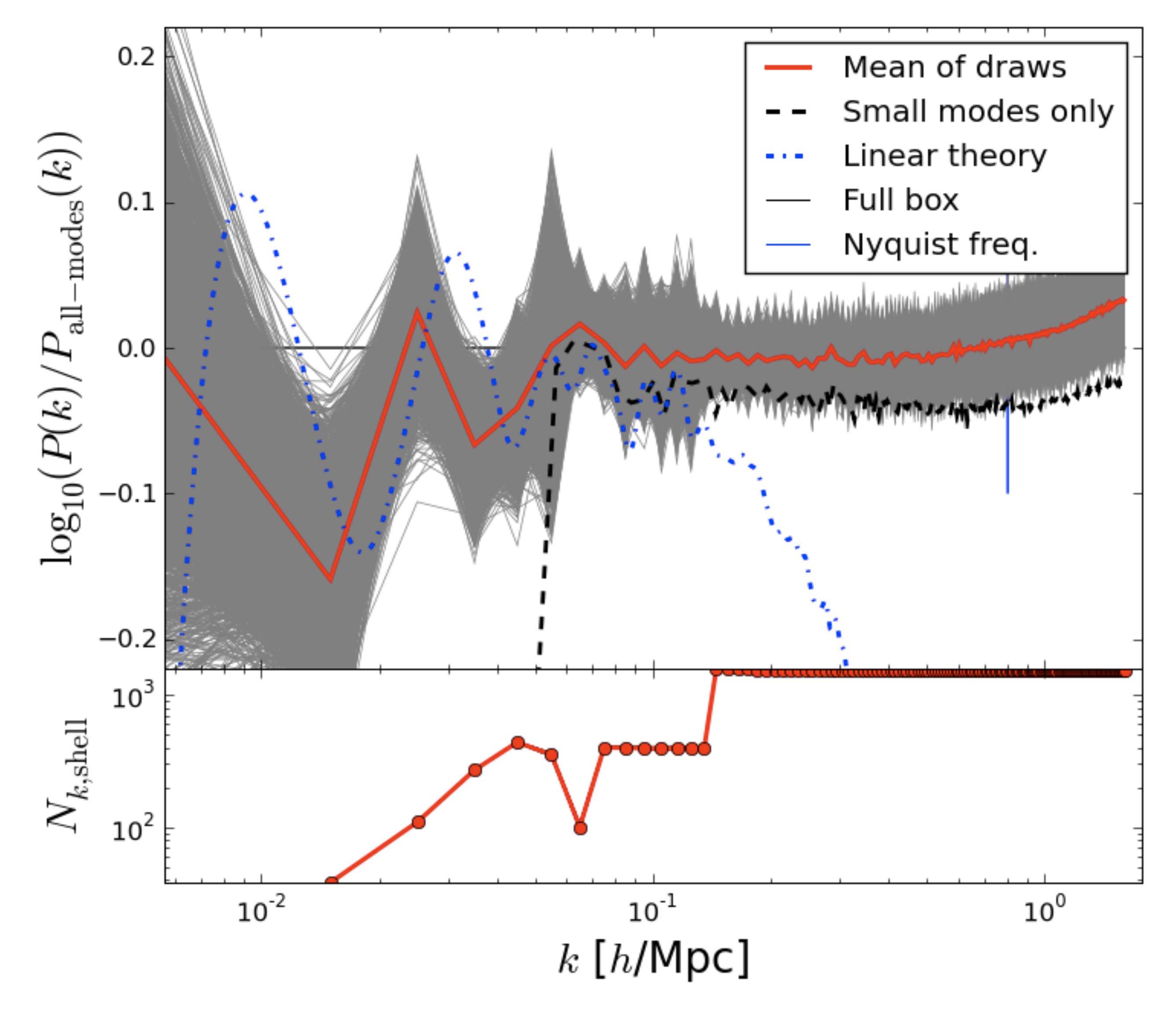}
	}
	\centerline{
	\includegraphics[scale=0.3]{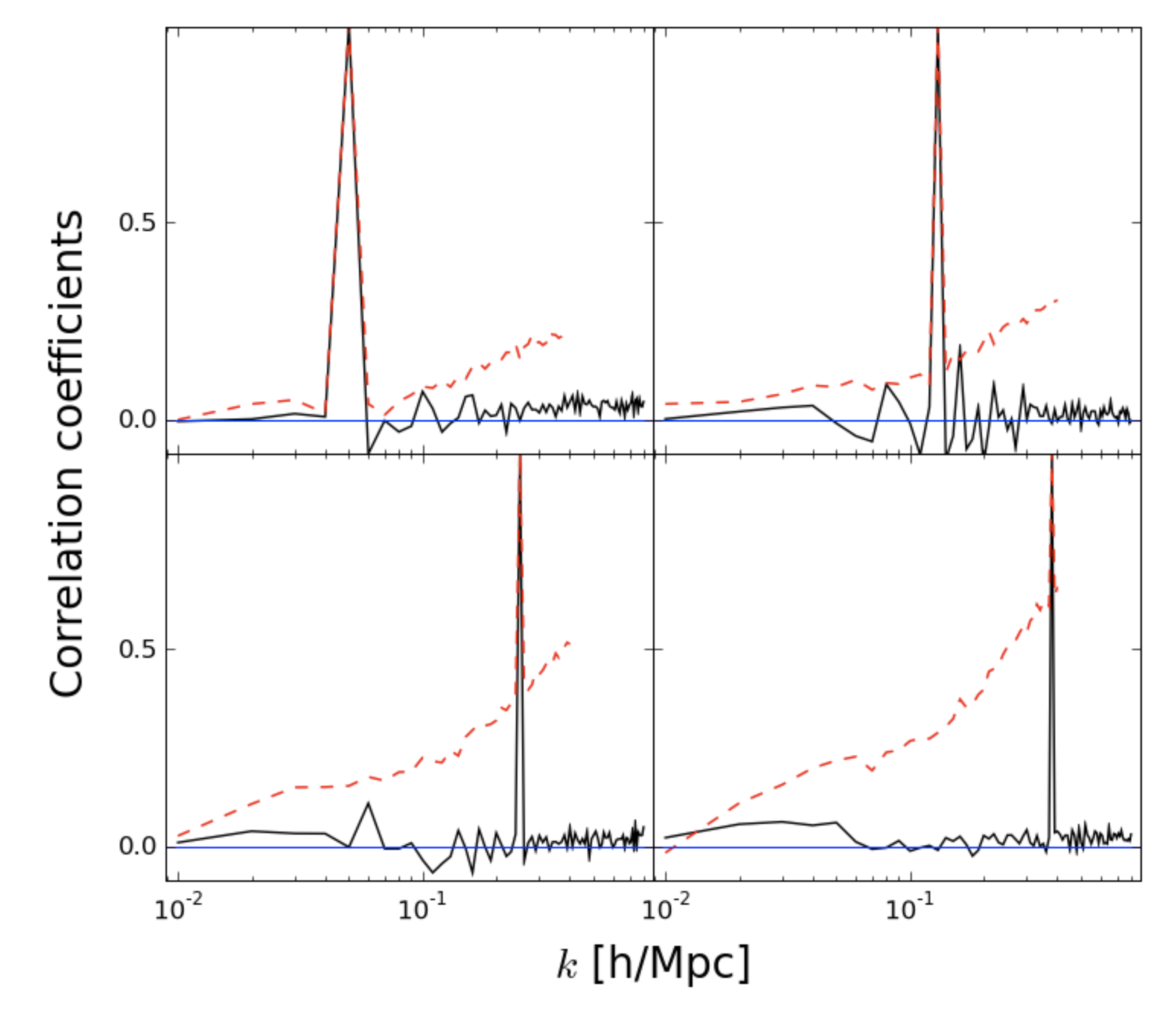}
	}
	\caption{\label{fg:distributedpowerspectrum}Top: Power spectrum estimates from 
	using only a sub-sample of the Fourier modes in each wavevector shell.  
	The power spectra are normalized by the power spectrum of the all-modes simulation 
	at $a=1$, so the wiggles at small $k$ are indicating the sample variance in the 
	all-modes simulation.
	Each power spectrum estimate from sub-sampled $k$-shells 
	is shown by the grey lines in the top panel.  
	The bottom panel of the top figure shows the number of Fourier modes used 
	to estimate the shell-averaged power spectrum.  The large jumps are where 
	we changed the number (and size) of the pixels used to sub-sample each $k$-shell.
	Bottom: Correlation coefficients of the power spectrum estimates (solid, black) compared 
	to the result from \citet{takahashi09} (dashed, red).}
\end{figure}

We applied this 
sub-sampling to a small-modes simulation with 500 large-scale mode resamplings.
The resulting power spectrum estimates and correlation coefficients of the 
estimated power spectrum covariance are shown in Figure~\ref{fg:distributedpowerspectrum}.
While the mean power spectrum is successfully recovered with this algorithm, 
the power spectrum covariance estimator has nearly zero off-diagonal terms, in stark 
contrast to the result from \citet{takahashi09} (shown in red, dashed lines) for wavenumbers 
$\gtrsim 0.1$~$h$Mpc$^{-1}$.  
One explanation for this could be that our method of sub-sampling modes in the 
wavevector shells is not actually selecting representative samples of all the 
four-point configurations in each shell, which would be fixed by a more precise sub-sampling
procedure.  

Because pursuing this algorithm further would significantly increase the computational 
complexity while limiting the range of applications, we instead turned to 
using multiple small-modes simulations with large-scale mode resampling 
to further explore the covariance matrix estimates as described in the next Section.

\subsubsection{Covariance matrix estimates from mode-resampling}
\label{sec:covestimation}
\begin{figure*}
  \centerline{
    \includegraphics[scale=0.5]{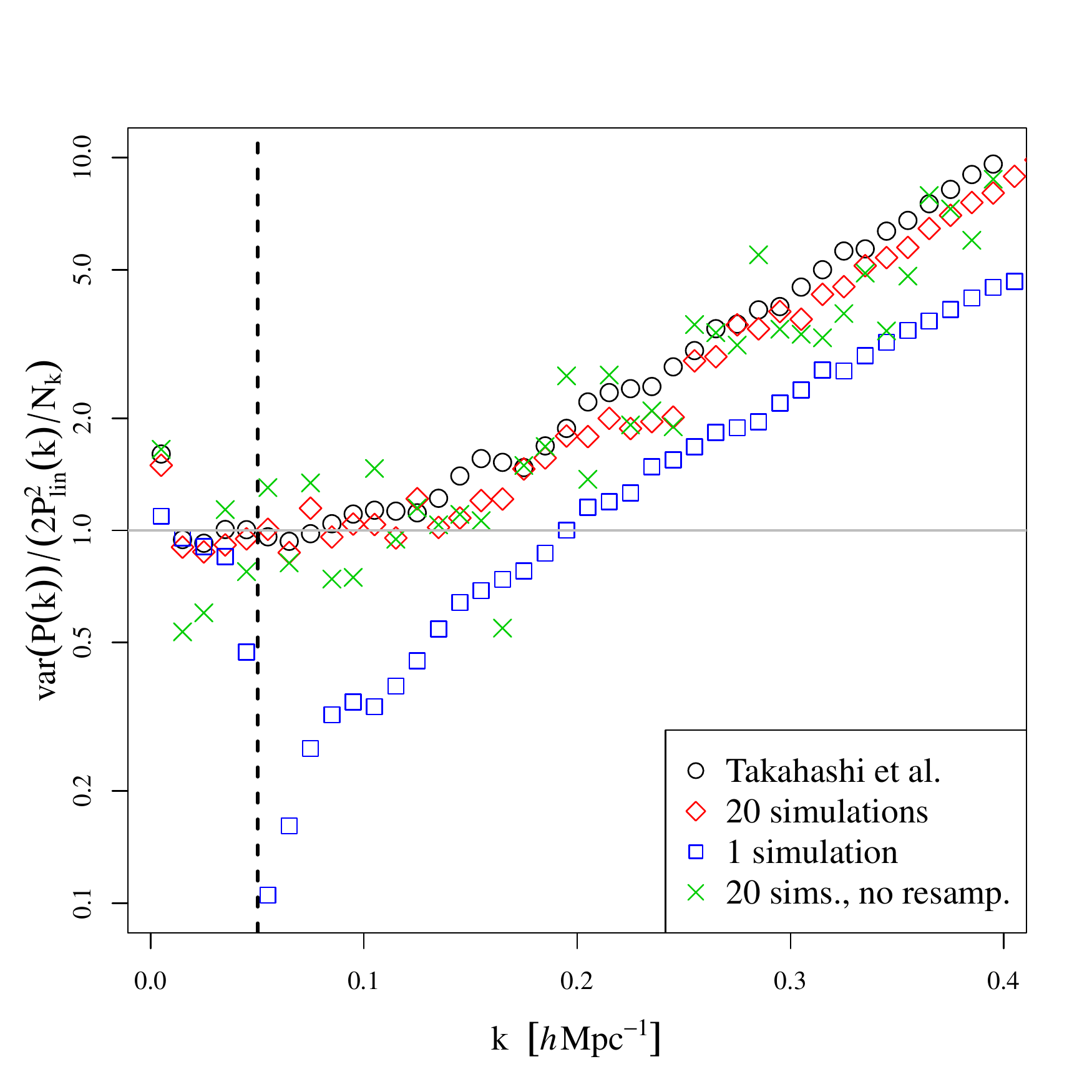}
  % }
  % \centerline{
    \includegraphics[scale=0.5]{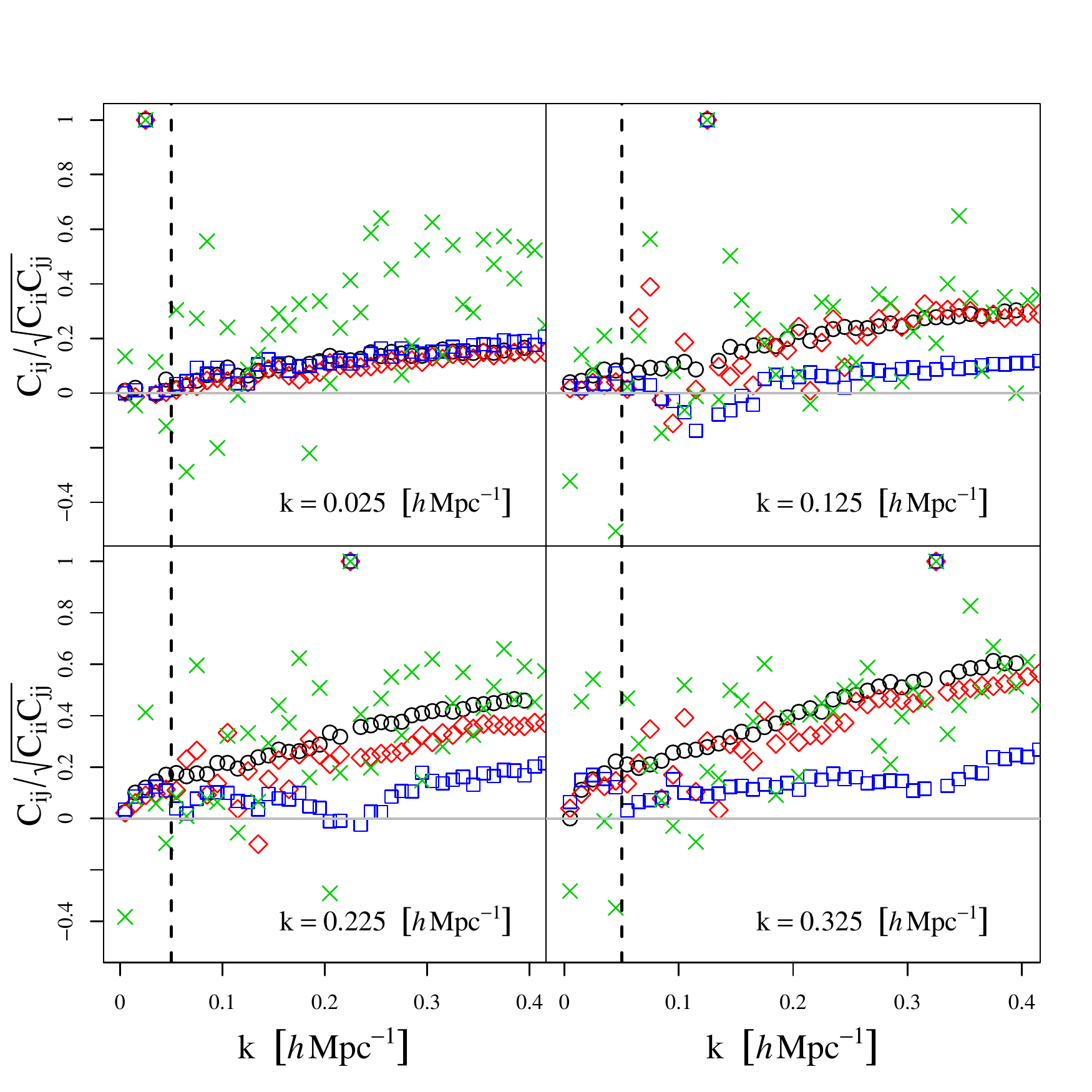}
  }
  \caption{\label{fg:covcomparison} Left: Power spectrum variance
  normalized to that for a Gaussian density field with the same linear power.  
  Right: Select correlation coefficients of the power spectrum covariance.  
 	Each panel corresponds to a row of the matrix of correlation coefficients, with 
  the row labeled by the annotated $k$ value in each panel (which is also where the 
  correlation coefficients are all equal to one each panel).
  The vertical dashed lines show the value of $\kthresh \equiv 8 k_F$.  The colors and 
  point styles are the same for both panels.  The black circles are values from 
	\citet{takahashi09}.  The red diamonds are derived from the sample covariance estimate 
	from 500 resamplings of each of 20 small-modes simulations.  The blue squares are the 
	estimates using 500 resamplings of only 1 small-modes simulation.  Finally, the green 
	$x$'s are the estimates when using 20 simulations without any resamplings (i.e., the 
	sample covariance estimate that would be derived without knowledge of the algorithm 
	in this paper.)}
\end{figure*}

The power spectrum variance and 
correlation coefficients estimated from 500 resamplings of the large-scale modes 
in a single small-modes simulation are shown by the blue squares in 
Figure~\ref{fg:covcomparison}.
Note that we use the same binning in wavenumber in all the cases shown as~\citet{takahashi09} 
in order to make an accurate comparison.
While the small-scale variance estimate is boosted beyond the Gaussian prediction, 
it is systematically biased low relative to the results of~\citet{takahashi09}.  
At intermediate wavenumbers just larger than $\kthresh$,
 the estimated variance is only $\sim20\%$ of the true value.  This is not 
 too surprising because it is at these scales where our algorithm for adding 
 large-scale modes is expected to be most inaccurate.  
 This is because our approximation that longer wavelength modes simply change 
 the effective local matter density must be wrong when the longer-wavelengths 
 are nearly equal to the smaller wavelengths to be perturbed.  The modes just 
 larger and smaller than $\kthresh$ may both contribute to coherent structures in the 
 mass density distribution so that resampling only some of the Fourier modes in these
 structures would severely underestimate the true variance on these scales.
 Also, we should note that the modes with wavenumbers just smaller than $\kthresh$ 
 are likely not strictly Gaussian distributed in the all-modes simulation, so that their true 
 distribution and coupling to small modes is not properly represented. 
 
 The right panel in Figure~\ref{fg:covcomparison} shows four rows of the matrix of 
 estimated correlation coefficients of the power spectrum.
 It is significant that the mode resampling generates nonzero correlation coefficients 
 between the large and small scales (blue crosses -- lower panels especially), 
 which are identically zero in the 
 unperturbed small-modes simulation.  However, the correlation coefficients 
 are again biased low by multiplicative factors of two or more. 
 
 We have checked that the biases in the estimated power spectrum variance 
 and correlation coefficients do not change when using more than 500 
 large-scale mode realizations.  Instead we speculate that this bias is real 
 and indicates that (at least for fixed $\kthresh$) the mode resampling algorithm 
 generates realizations of the matter density from only a subset of the full 
 nonlinear density probability distribution.  This makes sense intuitively 
 if we realize that the mode resampling as implemented here can never 
 generate a new set of halos in the simulation box, but 
 only moves and merges or de-merges the 
 halos already present in the small-modes simulation.
 Several previous 
 investigations have shown that the phase correlations in the Fourier 
 modes of the density field in a dark
 matter simulation are dominated by the masses and positions of the most massive 
 halos~\citep{coles03,neyrinck07}.  It makes sense then that the power spectrum variance on 
 small-scales would  also be quite sensitive to the particular realization of the most massive halos.  
 However, with a larger simulation volume the covariance estimates might be less 
 dependent on the particular realization of the massive halos because many different 
 halo configurations would be present.

If our mode resampling algorithm is simply sampling a subset of the density probability 
distribution, then the bias in the power spectrum variance and correlation coefficients
should decrease if we use more than one simulation.  
To check this hypothesis we ran 19 more small-modes simulations (for 20 total)
with different random number seeds in the initial conditions and 
added 500 resamplings of the large-scale modes to each simulation (for a total of 10000 
realizations of the density field).  
The sample covariance estimates improve considerably as shown by the  
red diamonds in Figure~\ref{fg:covcomparison}.  The biases in both the variance and 
the correlation coefficients are much smaller over all scales.  In addition, 
the extremely low dip in the variance estimates at intermediate scales when using only 
one simulation has disappeared entirely with 20 simulations.

Finally, we show an estimate of the covariance matrix using 20 simulations without any 
resampling of large-scale modes with the green crosses in Figure~\ref{fg:covcomparison}.  With such a 
small number of simulations the statistical noise dominates in the estimator.   Comparing with the 
estimator with 500 resamplings of each of the $N$-body simulations shows how much our 
mode-addition algorithm reduces the statistical errors. 

\subsubsection{Convergence rate of covariance estimates}
\begin{figure*}
	\centerline{
	\includegraphics[scale=0.46]{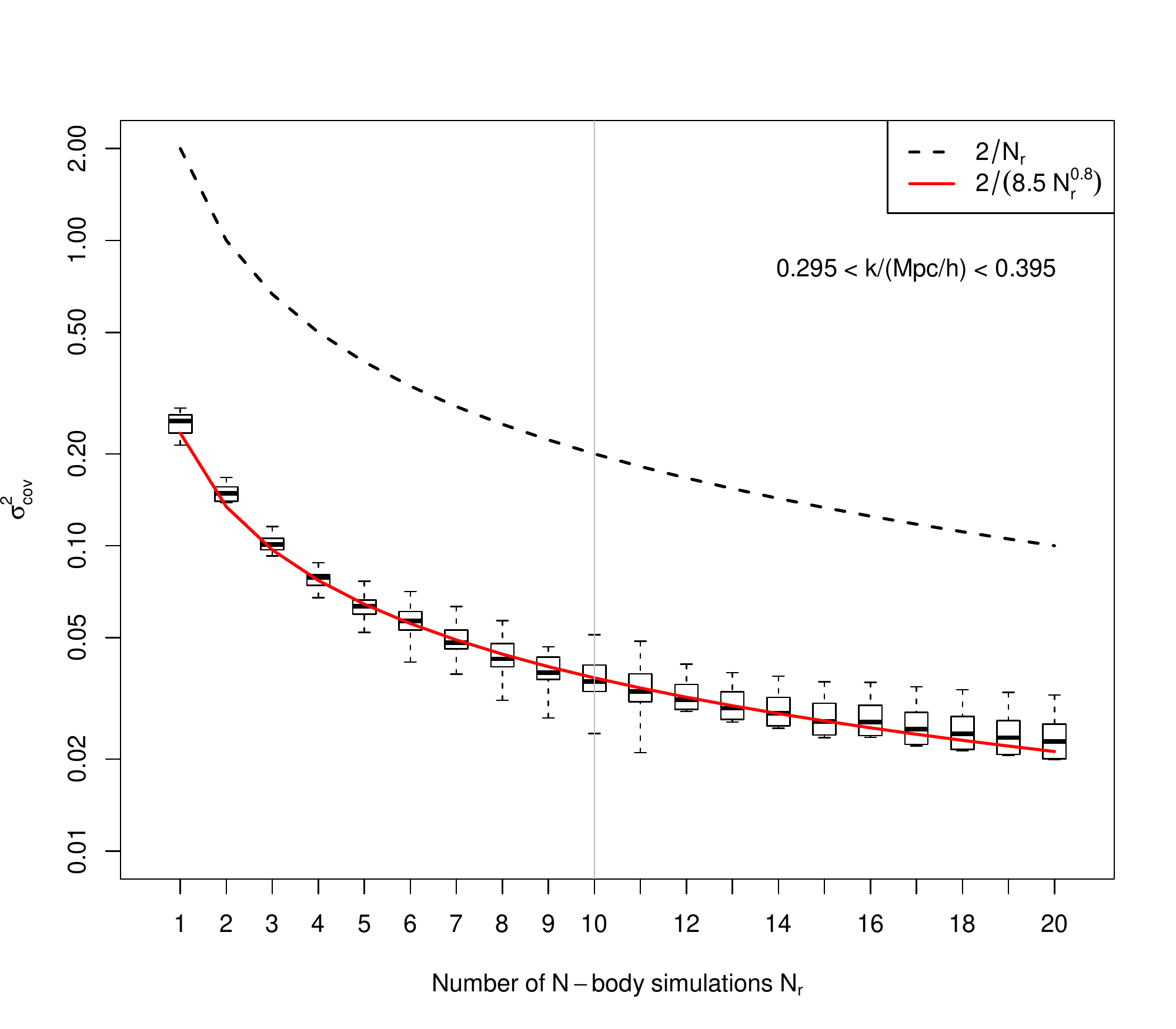}
	\includegraphics[scale=0.46]{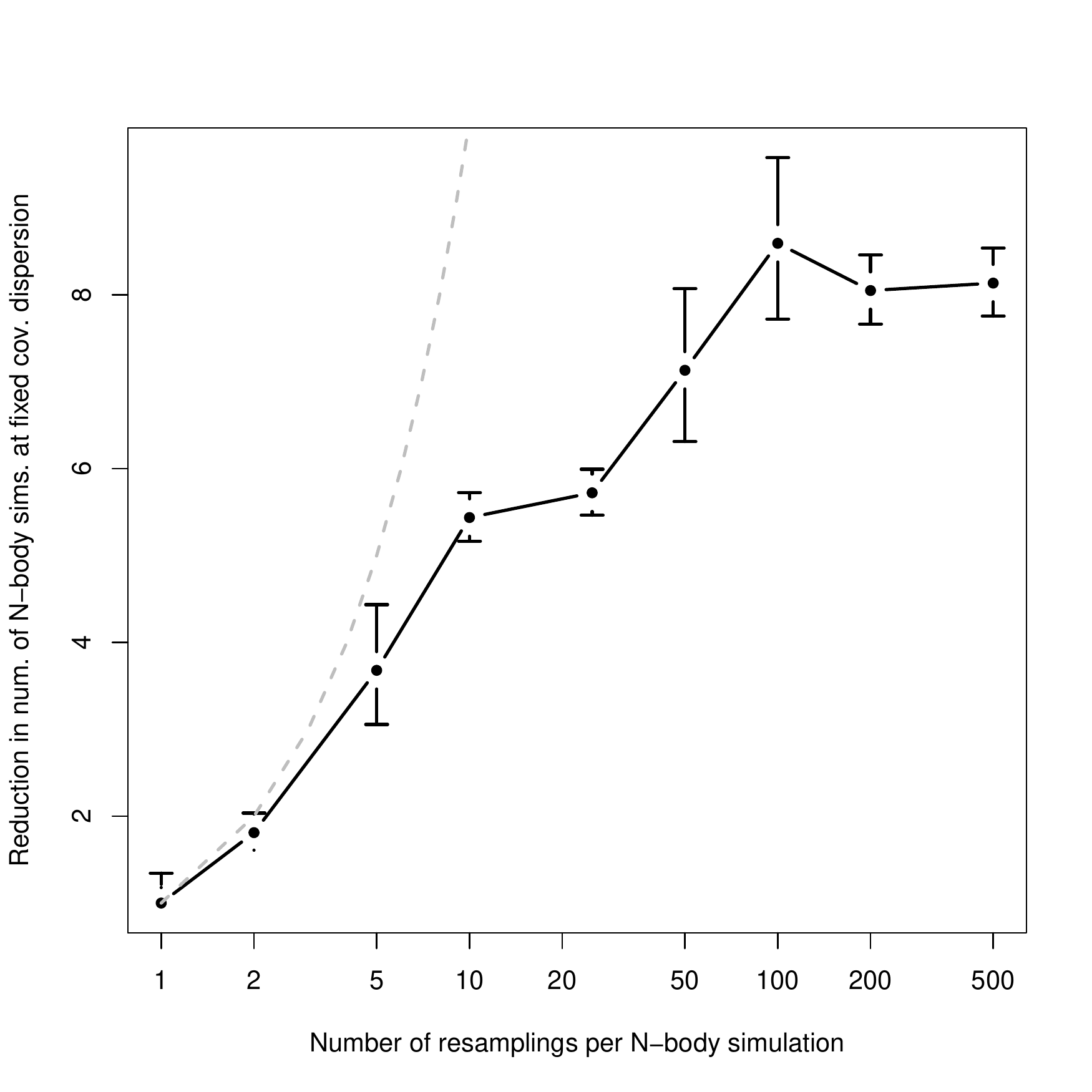}
	}
	\caption{\label{fg:convergence} Left: Dispersion in components of the power spectrum covariance matrix estimates as functions of the number of $N$-body simulations used to 
estimate the covariance.  The dispersion is measured relative to the result from~\citet{takahashi09} 
using 5000 simulations. Each $N$-body simulation has 500 resamplings of the large-scale modes, so a 
point on the plot has (number of $N$-body simualtions) $\times$ 500 total power spectra used to estimate the 
covariance.  The boxes indicate the medians (central black lines) and quartiles (top and bottom of the 
boxes) for the dispersion of the diagonal covariance elements within the wavenumber bins 
$0.295\le k/{\rm (h/Mpc)}\le 0.395$.  The dashed black line indicates the fit to the dispersion found 
in~\citet{takahashi09}.  The red line is a fit to the medians of our dispersion measurements.
Right: Normalization of the fit to $\sigma^2_{\rm cov}$ versus $N_r$ as a function of the number 
of mode resamplings per $N$-body simulation.  That is, we plot the fit parameter
$A\equiv 2/\left(\sigma^2_{\rm cov}N_{r}^b\right)$ just as for the red solid line in the left panel, 
but using between 1 and 500 mode resamplings per $N$-body simulation.  (We fit a different $b$ parameter 
for each value on the abcissa, but do not show these values.)
The error bars denote the 
95\% confidence intervals on the nonlinear least-squares parameter estimate.  The dashed grey line 
shows the expected scaling if each mode resampling was equivalent to running a new $N$-body simulation.
}
\end{figure*}

Although with 20 small-modes simulations the bias in the power spectrum covariance
estimate is much smaller than with only one simulation, a small bias is still noticeable 
in Figure~\ref{fg:covcomparison}.  In Figure~\ref{fg:convergence} we attempt to 
quantify the improvement in the covariance matrix estimates as more 
small-simulations are added by plotting the r.m.s. dispersion in elements of the 
power spectrum variance relative to the result from~\citet{takahashi09}.  
We show the distribution in the variance estimates for wavenumber bins with centers in 
$0.295 < k/({\rm Mpc}/h) < 0.395$ with the boxes in the left panel of Figure~\ref{fg:convergence}.  
The horizontal 
lines in the centers of the boxes denote the median of the variance estimates and the box top and 
bottom denote the first and third quartiles. 
We have selected only high-$k$ modes to plot in Figure~\ref{fg:convergence} because the 
variance estimate in this range has a roughly constant bias as seen in Figure~\ref{fg:covcomparison} 
and we are primarily interested in the accuracy of the covariance estimates on small scales where 
the deviation from the Gaussian model is most severe.  We show how the normalization of the fit 
to the covariance dispersion changes with the number of mode resamplings per $N$-body 
simulation in the right panel of Figure~\ref{fg:convergence}.  The reduction in the number of 
$N$-body simulations needed to give a certain error in the estimate covariance flattens 
around 100 resamplings per simulation.

\citet{takahashi09} performed a similar analysis and found that the dispersion in their 
sample variance estimators was well fit by $\sigma = 2 /N_{r}$, where $N_{r}$ is the 
number of $N$-body simulations (or realizations) used to compute the estimator (shown 
as the dashed black line in Figure~\ref{fg:convergence}).
We find a nonlinear least squares fit to our variance estimates of $2/(8.5 N_{r}^{0.8})$, 
which is shown by the solid red line in Figure~\ref{fg:convergence}.  From this we conclude 
that the sample variance estimator using resampled large-scale modes can achieve 
similar accuracy as the standard sample variance estimator but with $\sim 8$ times 
fewer $N$-body simulations.  However, it is difficult to extrapolate this convergence rate 
far beyond the 20 simulations we have actually performed because small changes in 
our choice of wavenumber bins or in our least-squares fit can cause large 
extrapolation errors.

%%%%%%%%%%%%%%%%%%%%%%%%%%%%%%%%%%%%%%%%%%%%%%%%%
\section{Discussion and Conclusions}
\label{sec:conclusions}
We have demonstrated an algorithm to resample the large-scale Fourier
modes of the density 
in an $N$-body simulation that successfully reproduces the nonlinear power spectrum
and significantly reduces the number of simulations required to estimate 
the power spectrum covariance matrix.  We expect our algorithm to aid 
the calculation of uncertainties for observational probes of the matter power spectrum 
such as weak lensing and galaxy clustering.  Because we can quickly provide multiple 
realizations of the density field, it is possible to estimate uncertainties after applying 
appropriate survey windows to the density, which can have significant impacts 
on the estimated power spectrum covariances.  This could be a distinct advantage over 
approaches that either precompute the power spectrum covariance without 
considering the survey window~\citep{semboloni07, takahashi09, sato09}, 
or use approximate or analytic methods that 
may not include the survey windows correctly~\citep[e.g.][]{crocce10}.  
We have no reason to 
believe the extension to redshift space will pose any particular challenges.
However, note that we have not yet tested the perturbation of the particle velocities so our 
results are limited to real-space statistics for now.

Our algorithm uses both Zel'dovich displacements as well as time-perturbations 
to match the effective local linear growth in over- or overdense regions.  
It is intriguing that neither of these operations alone is sufficient to 
reconstruct even the nonlinear power spectrum.  Evaluating the time-perturbation 
at the particle positions prior to the Zel'dovich move gives a bad result, 
indicating that the early-time movement of the particles is significant in 
determining the later growth of structures.  Any extensions of our algorithm 
will likely have to consider both of these components carefully.

Because applying the time-perturbation to remove large-scale modes from an
existing simulation generally requires many simulation snapshots far into the 
future, we found it onerous (or even impossible if resampling too many modes in 
a box that is too small)
to use existing simulations for mode-resampling.  
However, if the simulation volume is at least several Gpc on a side, then 
according to Figure~\ref{fg:growthtableinverse} it may be possible to subtract a 
limited number of large-scale modes with a feasible amount of computation.
For simulations volumes smaller than many Gpc on a side, 
we advocate running a new simulation with large-scale modes removed
in the initial conditions, which can then have new large-scale modes added 
directly in later snapshots using our algorithm.  Note that even for adding 
new large-scale modes the time-perturbation requires many snapshots 
closely spaced in scale factor both in the past and future of the desired 
time (but the simulation does not have to be run as far into the future as is 
needed for removing large-scale modes).  This means that application of 
our algorithm for, e.g. parameter estimation from a galaxy survey, will 
require special-purpose $N$-body simulations that will save significant 
computation time for the error analysis. 
% but may not be as suitable for 
% other types of analyses (at least the suitably would have to be tested
% first).  

% In particular, we hope to extend our algorithm to 
Our algorithm has other applications such as
including scaling 
of halo masses and concentrations so that mock galaxy catalogues could be 
constructed from each of the resampled density fields.  In our view, this is the 
only viable method to obtain robust uncertainties for estimating cosmological 
parameters from galaxy clustering and BAOs.  Our algorithm may even 
provide a feasible method to go beyond the work of~\citet{sefusatti06} for 
estimating the covariance of higher-order correlations.  
For parameter estimation 
with weak lensing, our resampled density fields could be directly input 
to existing ray-tracing pipelines.  However, one would need resampled 
densities for each of the lens planes (which could extend to $z\sim3$ for 
upcoming survey requirements).  This would require saving even more 
closely spaced snapshots of the small-modes simulation(s).  

The sample covariance matrix estimated from a large number of resamplings 
of a single simulation has nonzero correlations between large and small scales 
that are introduced by our mode addition algorithm.  But, the covariance elements
are biased low relative to the benchmark result from~\citet{takahashi09}, which 
used 5000 $N$-body simulations.  A possible explanation for this bias is that our 
method of resampling large-scale modes samples only a subset of the full 
probability distribution for the nonlinear density field.  We have shown that the bias
decreases monotonically as a function of the number of
$N$-body simulations used to estimate the covariance, which is consistent 
with this explanation.  If our mode-resampling algorithm were perfect at adding 
large-scale fluctuations, the bias in the covariance that we find 
would indicate that the distribution of the 
small-scale nonlinear density field is not entirely determined by the coupling 
with large-scale modes.  Rather, the phases of the small-scale Fourier modes in 
the initial conditions would be important as well, which contradicts the conclusion 
of~\citet{little91} who found little impact on the final 
nonlinear density when the small-scale phases were randomized in the 
initial conditions.  This could be because the eye notices mostly the 
large-scale structures in a dotplot of the particle positions
but there can be significant small-scale deviations that are less noticeable 
(as shown by the ``Zel'dovich only'' lines in Figure~\ref{fg:addpowersnapshots} versus the 
``Zel'dovich only" lower left panel of Figure~\ref{fg:dotplots}).

One area for future investigation is to study the dependence of the bias in 
the power spectrum sample covariance (or other measures of the distribution of 
the nonlinear density) with $\kthresh$.  This would quantify the relative importance 
of different large-scale modes on the nonlinear density distribution.  This could 
be a promising way to better understand the ``loss'' of information about 
cosmology in the matter power spectrum on 
translinear scales~\citep{rimes06, neyrinck06, neyrinck07}.  
One could also quantify on which scales (if any) the density in a large simulation 
becomes ergodic (so that the distribution of the density in sub-volumes of 
the simulation is equivalent to the distribution of the density field in an ensemble 
of large simulations).  The fact that our small-scale covariance estimates have 
such a large bias for $\kthresh\sim 0.5\,h/$Mpc (and in a single simulation) suggests
that assuming ergodicity on these scales is a fallacy when defining estimators 
of, at least some, summary statistics of the density.

Another application of our algorithm could be for tiling smaller simulation 
volumes to obtain the (properly correlated) density field in volumes that are 
too large to simulate, as originally proposed by~\citet{cole97}.  This would probably 
be most effective if we run several of our ``small-modes'' simulations and then 
tile them together by resampling the modes both larger and smaller than the 
fundamental frequency of the simulation volume.  A related application is re-running 
resimulation studies (that take a sub-volume of high-resolution $N$-body simulation and 
add additional physics) with different large-scale mode realizations.  This could be 
a relatively fast way to obtain different realizations of the resimulations with only 
one high-resolution $N$-body simulation at hand.  

Finally, we hope to apply our algorithm for obtaining estimates of both the mean 
power spectrum and its covariance as needed in simulation emulator 
frameworks~\citep{heitmann09, schneider11}.  Because we can estimate the 
power spectrum covariance with an easily achievable number of simulations, 
it will now be feasible to build an emulator for the cosmological-parameter 
dependence of the covariance~\citep{schneider08}.  Note that this paper is 
complementary to~\citet{angulo10} who described how to rescale the 
outputs of a single $N$-body simulation to mimic the outputs with 
a different cosmology, but who did not consider the time-perturbations we 
use here.  In combination, our algorithm and that of~\citet{angulo10} 
may serve to drastically reduce the number of simulations needed to construct 
simulation emulators for cosmological analysis.

\section*{Acknowledgments}
We thank Adrian Jenkins for extensive technical advice on Gadget-2 
and setup of initial conditions, David Weinberg for pointing us to his
earlier related work, Alex Szalay for explanations of the growth of 
correlations in the Fourier phases of the density field, Yanchuan Cai 
for advice on applying perturbation theory to (an ultimately failed attempt to)
add large-scale modes to our simulations and 
Mark Neyrinck, Bhuvnesh Jain, and Ravi Sheth for 
useful conversations.
SMC acknowledges the support of a Leverhulme Research Fellowship.
Some of the calculations for this paper were performed on the ICC
Cosmology Machine, which is part of the DiRAC Facility jointly funded by STFC,
the Large Facilities Capital Fund of BIS, and Durham University.
This work performed under the auspices of the U.S. Department of Energy by Lawrence Livermore National Laboratory under Contract DE-AC52-07NA27344.

\bibliographystyle{apj}
\bibliography{covestimation}

\label{lastpage}

\end{document}